\documentclass[aps,prl,twocolumn,superscriptaddress,amsmath,amssymb]{revtex4-2}
\usepackage{graphicx,bm,mathtools}
\usepackage[colorlinks=true,linkcolor=blue,citecolor=blue,urlcolor=blue]{hyperref}
\usepackage{xcolor}
\usepackage[normalem]{ulem}
\usepackage{cancel}

\newcommand{\e}{\mathrm e}

\newcommand{\norm}[1]{\lVert#1\rVert}
\begin{document}

\title{Chaos without Local Spectral Instability in a Kicked Oscillator}
\author{V. Citro}
\affiliation{Department of Industrial Engineering, University of Salerno, Fisciano, Italy}
\affiliation{CNR - Nanotec, Rome Unit, P.le Aldo Moro 5, 00185 Rome, Italy}
\author{G. Petrillo}
\affiliation{CNRS, ENS de Lyon, LPENSL, UMR5672, 69342, Lyon cedex 07, France}
\author{D. Sornette}
\email[Corresponding author: ]{dsornette@ethz.ch}
\affiliation{Institute of Risk Analysis, Prediction and Management (Risks-X), Southern University of Science and Technology, Shenzhen 518055, China}

\begin{abstract}
A damped harmonic oscillator kicked by a pulsed periodic potential can be chaotic although every local Jacobian has eigenvalues inside the unit circle. Damped rotation between kicks lets non-normal shears of opposite curvature compound, so ordered products of contracting Jacobians amplify perturbations. We derive the exact joint spectral radius of the Jacobian family and an exact Lyapunov-exponent decomposition showing how this geometric pumping overcomes dissipation.
The proposed mechanism can be realized with laser-cooled trapped ions in a pulsed optical lattice.
\end{abstract}
\maketitle

Since the seminal work of Lyapunov~\cite{Lyapunov1892}, linear stability analysis has played a central role in the study of dynamical systems and in predicting their asymptotic behavior. In one dimension, this analysis is conclusive: if a smooth map has a derivative whose magnitude is strictly less than one everywhere, the distance between any two trajectories contracts monotonically, precluding sustained exponential separation of nearby initial conditions. In contrast,  a contracting Jacobian in higher dimensions does not necessarily preclude chaos. A non-normal Jacobian can stretch a perturbation along one direction despite having all its eigenvalues inside the unit circle. When the Jacobian changes along a trajectory, the \emph{order} of these stretched directions determines whether transient amplification is recycled or lost. 
This mechanism is closely related to nonmodal amplification in fluid flows~\cite{Trefethen1993} and has also been explored as a route to chaos \cite{Sornette2026PRE,Sornette2026arXiv}. Existing examples are either piecewise smooth, relying on state-space partitions or modulo reinjection~\cite{Sornette2026PRE}, or smooth but not uniformly spectrally contracting, as in the optical and H\'enon-type maps of Ref.~\cite{Sornette2026arXiv}.

Here, we investigate a non-normal route to chaos in a simple, physically realizable system: a damped harmonic oscillator driven by a pulsed periodic potential. Although the eigenvalues of its local Jacobian remain strictly inside the unit circle throughout phase space, the system can sustain chaotic dynamics. This finding connects to a line of mathematical work showing that local spectral contraction does not ensure global attraction. Counterexamples to the discrete Markus--Yamabe conjecture established this failure of global stability~\cite{CimaGasullManosas1999}, and subsequent constructions demonstrated that smooth planar systems satisfying the same spectral constraints can nevertheless possess horseshoes~\cite{AddasGomes2011}. While these results establish the mathematical possibility of chaos under local spectral contraction, they do not by themselves identify the dynamical mechanism that generates it in a physically realizable system. In this letter, we address this question by providing a mechanical realization of the non-normal route to chaos, deriving exact amplification criteria for compositions of locally stable Jacobians, and quantifying how state-dependent geometric amplification can overcome dissipation along nonlinear trajectories.

\paragraph*{A pulsed oscillator.}
Consider a particle of mass $m$ confined by a harmonic trap centered at $x_0$, with natural frequency $\omega_0$ and damping rate $\gamma<\omega_0$. A sinusoidal substrate acts on it in short pulses separated by $T$:
\begin{equation}\label{eq:eom}
\ddot x+2\gamma\dot x+\omega_0^2(x-x_0)
=\frac{F_0\tau}{m}\sin(kx)\sum_{n\in\mathbb Z}\delta(t-nT).
\end{equation}
Here $\tau$ is the effective pulse duration and $2\pi/k$ the substrate wavelength. The spring and corrugation evoke the Prandtl--Tomlinson description of atomic friction~\cite{Prandtl1928,Tomlinson1929,Vanossi2013}, and a trapped ion illuminated by a pulsed standing wave is one possible realization~\cite{Bylinskii2015}. 

Between successive kicks, the oscillator undergoes a damped rotation in phase space. To describe its dynamics from one kick to the next, we introduce the rescaled variables and parameters $X=k(x-x_0)$, $P=(\dot X+\gamma X)/\omega$
where $\omega=(\omega_0^2-\gamma^2)^{1/2}$ is the damped oscillation frequency. We also define the damping factor 
$\lambda=\e^{-\gamma T}$, the rotation angle $\theta=\omega T$, the dimensionless kick strength $K=kF_0\tau/(m\omega)$, and the phase $\varphi=kx_0$. Sampling the state immediately before each kick then yields the exact stroboscopic map

\begin{equation}\label{eq:physical}
\binom{X_{n+1}}{P_{n+1}}=
\lambda R(\theta)\binom{X_n}{P_n+K\sin(X_n+\varphi)},
\end{equation}
where $R(\theta)$ is the planar rotation with rows $(\cos\theta,\sin\theta)$ and $(-\sin\theta,\cos\theta)$. Its Jacobian, $\lambda R(\theta)(I+K\cos(X_n+\varphi)E)$ with $E_{21}=1$ and $E^2=0$, combines an attenuation, a rotation, and a state-dependent non-normal shear. For kicks a quarter of the damped period apart, $\theta=\pi/2$, elimination of $P_n$ gives the two-dimensional recurrence
\begin{equation}\label{eq:rec}
\begin{split}
y_{n+1}=&\mu\sin y_n-b y_{n-1}+C, \quad y_n=X_n+\varphi=kx_n\\
b=&\lambda^2 ,\quad \mu=\lambda K,\quad C=(1+b)\varphi ~.
\end{split}
\end{equation}
We use $(b,\mu,C)=(0.9,1.7076,1.885)$ unless stated otherwise.

\paragraph*{Every local Jacobian is spectrally stable.}
For the state $z_n=(y_{n-1},y_n)$, the Jacobian is
\begin{equation}\label{eq:J}
J(t_n)=\begin{pmatrix}0&1\\-b&t_n\end{pmatrix},
\qquad t_n=\mu\cos y_n,
\end{equation}
with trace $t_n$ and determinant $b$. The condition $0<b<1$ and $\mu<2\sqrt b$ makes the discriminant $t_n^2-4b$ negative \emph{at every point of $\mathbb R^2$}. Consequently
\begin{equation}\label{eq:local}
\rho(J(t_n))=\sqrt b<1\qquad\text{globally}.
\end{equation}

Therefore, the possible presence of chaotic dynamics cannot be attributed to a hidden local eigenvalue crossing: all local Jacobian eigenvalues remain strictly inside the unit circle throughout phase space, including outside the attractor. The map is analytic, invertible, and globally bounded forward in time (Supplemental Material, SM). At the working parameters, the system has a unique stable fixed point and no period-two orbit, showing that the presence of chaotic dynamics is not accompanied by these simpler forms of recurrent motion.

Local spectral contraction persists beyond quarter-period kicking.
For a general phase $0<\theta<\pi$, all local Jacobians have
complex-conjugate eigenvalues of modulus $\lambda<1$ provided
$K/2<\tan(\theta/2)<2/K$.
At our working value $K=1.8$, this permits a detuning of approximately
$\pm 6^\circ$ around $\theta=\pi/2$
($83.97^\circ<\theta<96.03^\circ$).
Because every inequality in our computer-assisted proof holds with a finite margin, the horseshoe survives small detuning (SM, Sec.~S10); whether the chaotic attractor persists across the whole window remains open and can be tested experimentally.

\begin{figure}[t]
\centering
\includegraphics[width=\columnwidth]{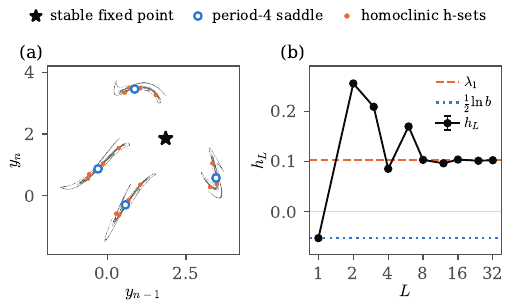}
\caption{\label{fig:attractor}(a) Numerically observed strange attractor (gray) of Eq.~\eqref{eq:rec} at $(b,\mu,C)=(0.9,1.7076,1.885)$, coexisting with a stable fixed point (star). Blue circles identify the period-four saddle, and orange points mark the centers of the covering sets ($h$-sets) along the homoclinic excursion used in the computer-assisted proof. (b) Ordered-product spectral rate $h_L$ from Eq.~\eqref{eq:hL}: despite $h_1=\tfrac12\ln b<0$, $h_2\simeq0.255$ and $h_L$ approaches the measured $\lambda_1\simeq0.103$ (error bars: $95\%$ confidence intervals over ten independent orbits).}
\end{figure}

\paragraph*{How contractions make an expanding product.}
The chaotic attractor in Fig.~\ref{fig:attractor}(a) is characterized by a maximal Lyapunov exponent of $\lambda_1=0.1030\pm0.0002$, with the uncertainty denoting a 95\% confidence interval over ten independent orbits of $10^6$ iterations each. Since $\det J=b$, the second Lyapunov exponent follows as $\lambda_2=\ln b-\lambda_1\simeq-0.208$, yielding a Kaplan–Yorke dimension of approximately $1.49$. The persistence of chaotic dynamics is further supported by 64 trajectories, each followed for $3\times10^7$ iterations, all of which remain in the chaotic regime (SM, Sec. S7).

The phase space also contains distinct attracting regimes. Sampling an $81\times81$ grid of initial conditions over the square $[-4,5]^2$ in the $(y_{n-1},y_n)$ plane, we find that $44.6\pm0.6\%$ of initial conditions converge to the stable fixed point, $36.9\pm0.6\%$ to the chaotic attractor, and $18.4\pm0.5\%$ to other, all-periodic attractors. The coexistence of these basins shows that the system's response depends on the initial condition: sufficiently small perturbations near the stable fixed point decay, whereas finite perturbations can drive the system into a different attracting regime.                                                                                                                                                                                                                                                                                                                                                                                                                                                                                                                                                                                                                                                                                                                                                                                                                                                                                                                                                                                                                                                                                                                                                                                                                                                                                                                                                                                                                                                                                                                                                                                                                                                                                                                                                                                                                                                                                                                                                                                                                                        

The one-step spectral rate is constant, $h_1=\frac12\ln b\simeq-0.0527$, on \emph{every} orbit. In contrast, the finite-length diagnostic introduced in Ref.~\cite{Sornette2026arXiv},
\begin{equation}\label{eq:hL}
h_L=\frac1L\left\langle\ln\rho(J_{n+L-1}\cdots J_n)\right\rangle,
\end{equation}
measures the growth produced by $L$ successive tangent maps. As shown in Fig.~\ref{fig:attractor}(b) (see also SM, Sec.~S7), it is already positive at $L=2$, where $h_2=0.2554\pm0.0001$, and approaches $\lambda_1\simeq0.103$ at longer lengths. At the stable fixed point, $h_L=\frac12\ln b <0$ for \emph{every} $L$. The two invariant states therefore have \emph{identical one-step spectral information} but different geometric composition.

The distinction is particularly clear in the singular-vector representation $J_n=U_n\Sigma_nV_n^{\mathsf T}$. Two consecutive kicks produce
\begin{equation}\label{eq:SVD}
J_{n+1}J_n=U_{n+1}\Sigma_{n+1}
\underbrace{(V_{n+1}^{\mathsf T}U_n)}_{\mathcal C_n}
\Sigma_nV_n^{\mathsf T}.
\end{equation}
The singular values in $\Sigma_n$ measure the stretching available at each kick, whereas $\mathcal C_n$ measures how the next kick receives the direction just stretched. The gains cancel when successive amplification axes are poorly aligned and compound when they cooperate. A fixed Jacobian can have a singular value larger than one and yet decay under repeated application \cite{Sornette2026arXiv}, while the changing curvatures of the oscillator keep rebuilding the amplifying geometry. 

A two-step product has trace $t_nt_{n+1}-2b$ and determinant $b^2$. Throughout the globally complex-spectrum regime, 
It is spectrally expanding, $\rho(J_{n+1}J_n)>1$, precisely when
\begin{equation}\label{eq:two}
t_nt_{n+1}<-(1-b)^2.
\end{equation}
One kick shears momentum, and a quarter-period rotation turns this into a position shear. A sufficiently strong shear of opposite sign then creates a hyperbolic \emph{two-kick} product. On the chaotic orbit, $50.0\%$ of successive pairs visit opposite substrate curvatures and $49.8\%$ satisfy Eq.~\eqref{eq:two}, while at the stable fixed point the curvature never changes sign. Longer products can expand even when none of their consecutive two-step products does, and even without opposite-sign curvatures (SM, Sec.~S5).
 
To isolate the role of the temporal ordering of successive
Jacobians, we perform a numerical experiment in which we
retain the same $N=2\times10^5$ Jacobians $J_n$ sampled
along the chaotic attractor, but change the order in which
they are multiplied. For each ordering, we calculate the
asymptotic logarithmic growth rate per iteration of an
infinitesimal perturbation, estimated as
$N^{-1}\ln\|J_{N-1}\cdots J_0\|$, with the spectral norm, for the corresponding
reordered product. In the original chronological order, this rate reproduces the maximal Lyapunov exponent, $0.1027\pm0.0006$. Randomly permuting the same Jacobians increases the growth rate to $0.1857\pm0.0010$, indicating that random sequences
can exploit the available non-normal amplification \cite{TroudeSornette2026} even
more efficiently than the trajectory generated by the
nonlinear dynamics. Conversely, sorting the Jacobians
according to their local curvature parameter
$t_n=\mu\cos y_n$ reduces the growth rate to $-0.0527$, essentially the uniform local spectral contraction rate $\frac12\ln b$, because sorting places matrices with similar curvatures next to one another and suppresses the repeated alternation between restoring and anti-restoring curvatures that favors amplification. All three experiments use exactly the same Jacobian matrices, preserving their eigenvalues, singular values, and entire distribution of local curvatures, and only their temporal ordering changes. The differences between the three growth rates, about two orders of magnitude larger than their statistical uncertainties, thus demonstrate that sustained amplification is governed by the geometric composition of successive non-normal transformations. The finite-length spectral diagnostics, their dependence on the ordering of successive Jacobians, their uncertainties, and additional numerical results are presented in SM, Sec.~S7.

\paragraph*{An exact amplification bound.}
The two-kick mechanism can be extended to a theorem covering products of \emph{any} length. Let $\mathcal J_{b,\mu}=\{J(t):|t|\le\mu\}$ and
let $\widehat\rho(\mathcal J_{b,\mu})$ denote its \emph{joint spectral radius} (JSR): the largest asymptotic per-step growth possible over arbitrary words of admissible Jacobians~\cite{RotaStrang1960,Jungers2009}. Since \(t_n=\mu\cos y_n\), every Jacobian encountered along a trajectory 
belongs to \(\mathcal J_{b,\mu}\). The Lyapunov exponent measures amplification only along the sequences that the nonlinear dynamics actually generates.
Let us define $\|(u,v)\|_b^2=bu^2+v^2$. Then, for every $L\ge1$,
\begin{align}\label{eq:jsr}
\max_{|t_1|,\ldots,|t_L|\le\mu}
\norm{J(t_L)\cdots J(t_1)}_b&=q^L,\nonumber\\
\widehat\rho(\mathcal J_{b,\mu})=q&=\frac{\mu+\sqrt{\mu^2+4b}}2.
\end{align}
These equalities follow from a common quadratic norm that bounds
all finite products. Introducing $S=\mathrm{diag}(\sqrt b,1)$ and $B(t)=SJ(t)S^{-1}$, the largest singular value of $B(t)$ is
$
\|B(t)\|_2
=\frac{|t|+\sqrt{t^2+4b}}{2}\le q,
$
so every product of $L$ admissible Jacobians satisfies $\|J(t_L)\cdots J(t_1)\|_b\le q^L$, since $\|v\|_b=\|Sv\|_2$.
This bound is attained by alternating the extreme curvatures
$t=\pm\mu$. Writing $B_+=B(\mu)$, the identity $B(-\mu)=-B_+^{\mathsf T}$ reduces alternating products of lengths $2k$ and $2k+1$ to $(-B_+B_+^{\mathsf T})^k$ and $(-1)^k B_+(B_+^{\mathsf T}B_+)^k$, respectively.
Their operator norms are exactly $q^{2k}$ and $q^{2k+1}$.
Thus, the maximum amplification in this norm is $q^L$
at every finite length, and the joint spectral radius is
$\widehat\rho(\mathcal J_{b,\mu})=q$.
The complete singular-value proof and the global-contraction corollary are given in SM, Sec.~S4.

The exact joint spectral radius establishes a sharp boundary
between global contraction and the possibility of geometric
amplification. For $\mu<1-b$, one has $q<1$, implying
that every admissible Jacobian contracts in the same
weighted norm. Consequently, the nonlinear map possesses
a unique globally attracting fixed point, a result that
holds for any differentiable kick force satisfying $|f'|\le\mu$, including the sinusoidal force considered here. When $\mu>1-b$, one has $q>1$: suitably ordered
products of individually contracting Jacobians can
amplify perturbations exponentially. The chaotic attractor at $b=0.9$ and $\mu=1.7076$ shows how much of this potential the dynamics actually uses. The largest
theoretically available growth rate per iteration is
$\ln q\simeq0.7562$, while the measured maximal Lyapunov
exponent is only $\lambda_1\simeq0.103$. The oscillator
therefore realizes only part of the amplification
potential permitted by its Jacobians. The condition $\mu>1-b$ is thus necessary for chaos: it opens the possibility of geometric amplification, which the nonlinear dynamics sustains only for suitable sequences of curvatures. Finally,
requiring all local eigenvalues to remain complex imposes
$\mu<2\sqrt b$. The corresponding regime of locally
contracting complex spectra with potentially expanding
products, $1-b<\mu<2\sqrt b$, exists only for
$b>3-2\sqrt2 \approx 0.1716$.

\begin{figure}[t]
\centering
\includegraphics[width=\columnwidth]{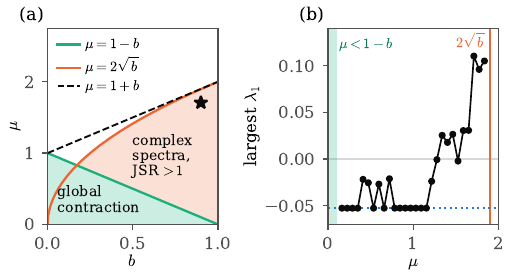}
\caption{\label{fig:phase}(a) Exact analytical boundaries: global contraction for $\mu<1-b$; locally contracting complex spectra but potentially expanding products for $1-b<\mu<2\sqrt b$; and possible local spectral instability beyond $\mu=1+b$. (b) Largest numerically measured maximal Lyapunov exponent versus $\mu$ at $b=0.9$, scanning the loading $C$ and initial conditions. Positive values identify numerically chaotic attracting regimes. Detailed panel interpretation, scanning protocol, and onset values are given in SM, Sec.~S3.}
\end{figure}

Figure~\ref{fig:phase}(a) summarizes the exact analytical boundaries.
For $\mu<1-b$, every admissible Jacobian product contracts,
and the oscillator converges globally to a unique fixed
point. For $1-b<\mu<2\sqrt b$, expanding products become
possible even though every individual Jacobian has complex
eigenvalues of modulus $\sqrt b<1$. In the larger interval
$2\sqrt b<\mu<1+b$, the eigenvalues can become real but
remain strictly inside the unit circle. These boundaries
characterize the amplification capabilities of the Jacobian
family, independently of the actual nonlinear trajectories.

Panel (b) shows the behavior of the sinusoidally kicked oscillator at $b=0.9$. The exact threshold for possible amplification is $\mu=1-b=0.1$, yet the numerical scan over the loading $C$ and initial conditions detects
nontrivial attracting states only above $\mu\simeq0.36$,
and chaotic attractors with a positive maximal Lyapunov
exponent only above $\mu\simeq1.26$. Thus, the possibility
of exponential amplification opens well before chaos
actually appears. The reason is that the nonlinear dynamics explores a restricted set of Jacobian sequences:
the successive potential curvatures are selected
self-consistently by the evolving state and the loading
$C$. Only when this dynamics generates sufficiently
favorable sequences of curvatures can the available
non-normal amplification overcome dissipation and sustain
chaos.

\paragraph*{Exact decomposition of the maximal Lyapunov exponent.}
We now present an exact decomposition of the maximal Lyapunov exponent 
that reveals how the geometric composition of successive Jacobians generates sufficient amplification to 
overcome damping along the chaotic trajectory.
Each \emph{frozen} Jacobian is contracting in its \emph{own} quadratic metric:
\begin{equation}\label{eq:metric}
G(t)=\begin{pmatrix}b&-t/2\\-t/2&1\end{pmatrix}>0,
~ J(t)^{\mathsf T}G(t)J(t)=bG(t).
\end{equation}
The shape of this contracting ellipse changes with the substrate curvature $t_n$. For an infinitesimal displacement $\xi_{n+1}=J(t_n)\xi_n$, set $E_n=\xi_n^{\mathsf T}G(t_n)\xi_n$ and write $\xi_n=(u_n,v_n)^{\mathsf T}$. Locally $J_n^{\mathsf T}G(t_n)J_n=bG(t_n)$, but the next step is assessed in the generally different metric $G(t_{n+1})$. Inserting $G(t_{n+1})-G(t_n)$ between two consecutive tangent steps gives the exact balance
\begin{equation}\label{eq:pump}
E_{n+1}=bE_n+W_n,~
W_n=-(t_{n+1}-t_n)u_{n+1}v_{n+1}.
\end{equation}
Since $G(t)$ is uniformly positive definite for $|t|\le\mu<2\sqrt b$, we may take logarithms and sum over successive iterations. All intermediate terms cancel, leaving the logarithmic growth of the perturbation between its initial and final states, and in the long-time limit we obtain the identity
\begin{equation}\label{eq:lambda}
\lambda_1=\tfrac12\ln b+
\tfrac12\left\langle\ln\left(1+\frac{W_n}{bE_n}\right)\right\rangle.
\end{equation}
The first term is uniform damping, and the second is cumulative \emph{geometric pumping} between incompatible local contraction metrics. Since the two metrics are uniformly equivalent to ordinary Euclidean length, the logarithmic sum of these mismatches contributes directly to the Lyapunov exponent, without a stochastic approximation or a small-curvature-change assumption. In this sense, the orbit rectifies transient amplification in tangent space. At the fixed point, $G$ is constant and $W_n=0$. On the chaotic attractor, the pumping term equals $0.1557\pm0.0002$ and overcomes $\frac12\ln b\simeq-0.0527$. Along ten independent orbits of $10^6$ iterations, Eq.~\eqref{eq:lambda} reproduces the growth rate of the tangent vector itself to within $1.3\times10^{-6}$, the bounded boundary term of the telescoping sum, and the Lyapunov exponent computed independently by QR reorthonormalization to within $6\times10^{-5}$.  The full derivation, a numerical evaluation procedure, and the extension to nonlinear memory are given in SM, Sec.~S6.

\paragraph*{A rigorous chaotic horseshoe and the onset of intermittency.}
To establish chaos beyond numerical Lyapunov estimates,
we construct a computer-assisted proof based on the
dynamics near a period-four saddle and a near-homoclinic
excursion. Interval arithmetic verifies 63 covering
relations, yielding two distinct symbolic itineraries
that generate a chaotic horseshoe for the 60th iterate.
This establishes positive topological entropy,
\begin{equation}\label{eq:entropy}
h_{\mathrm{top}}(F)\geq\frac{\ln 2}{60}>0,
\end{equation}
and thus rigorously demonstrates chaos despite the
globally contracting local spectra. The proof establishes
a compact chaotic invariant set, while the attracting
character of the numerically observed chaotic state
remains supported by simulations. 
The construction of the covering sets, their interval verification, and the resulting positive-entropy invariant measure are detailed in SM, Sec.~S8.

\begin{figure}[t]
\centering
\includegraphics[width=\columnwidth]{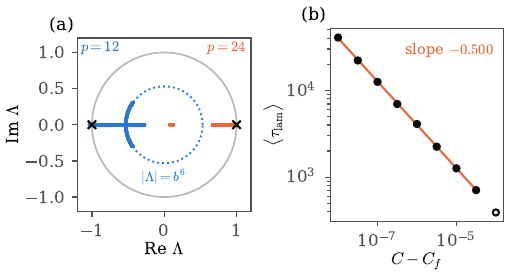}
\caption{\label{fig:route}(a) Floquet multipliers of a period-12 orbit (blue) move on the circle $|\Lambda|=b^6$ and collide on the negative real axis at $C^*=1.842847$, after which one of them crosses $-1$. The period-24 branch is orange. (b) Mean laminar residence time just beyond the saddle-node at $C_f\simeq1.84593003$ scales as $(C-C_f)^{-0.500\pm0.002}$, compatible with type-I intermittency, and is measured on the chaotic transient that precedes capture by a periodic orbit (SM, Sec.~S7). }
\end{figure}

The bifurcation sequence reveals how geometric amplification
develops into chaos (Fig.~\ref{fig:route}). At the same $b$ and $\mu$, for the lower loading $C=1.838$, the two Floquet multipliers of a period-12 orbit are complex,
with modulus $|\Lambda|=b^6=0.531441$. They collide on the
real axis at $C^*=1.842847$, after which one crosses $-1$
at $C\simeq1.844882$, producing a period-24 orbit.
This branch terminates at a saddle-node bifurcation at
$C_f\simeq1.8459300302$. Beyond this threshold, long
nearly periodic laminar phases alternate with chaotic bursts,
with a mean residence time scaling as
$\langle\tau_{\rm lam}\rangle\propto(C-C_f)^{-0.500\pm0.002}$ (statistical error, with a shift of the exponent $0.5$ of at most $0.013$ when the laminar threshold is halved or doubled),
consistent with type-I intermittency~\cite{PomeauManneville1980} (numerical details in SM, Sec.~S7). For $C-C_f<7.6\times10^{-5}$, this intermittent dynamics is a chaotic transient that ends on a stable periodic orbit of period $28$, $56$ or $112$, and sustained intermittent chaos sets in beyond that loading, where these orbits no longer exist. Since the determinant of the period-$p$ monodromy matrix is
$b^p$, its complex-conjugate Floquet multipliers necessarily
remain on a circle of radius $b^{p/2}$ in the complex plane. Their collision
and subsequent splitting along the real axis redistribute
the contraction between the two Floquet directions,
eventually allowing one multiplier to cross the unit circle.
Thus, the ordered composition of locally contracting,
elliptic Jacobians generates a collective Floquet instability
that develops through period doubling and a saddle-node
bifurcation into intermittent chaos.

\paragraph*{Discussion.}
We have shown that a damped, periodically kicked oscillator
can sustain chaos even though its local Jacobian spectrum
contracts everywhere. We derived exact amplification bounds
and a Lyapunov-exponent decomposition that explain how
successive non-normal shears overcome dissipation, and
established chaotic invariant dynamics through a
computer-assisted horseshoe proof. Laser-cooled ions in a
pulsed optical lattice offer a promising experimental
realization of this mechanism. In the SM, we provide a
broader physical interpretation, detailed experimental
implementation, mechanical-energy balance, and connection
to nonlinear feedback (Secs.~S1, S9--S11).

\clearpage
\onecolumngrid
\setcounter{page}{1}
\renewcommand{\thepage}{S\arabic{page}}
\setcounter{equation}{0}
\renewcommand{\theequation}{S\arabic{equation}}
\renewcommand{\theHequation}{SM.\arabic{equation}}
\setcounter{figure}{0}
\renewcommand{\thefigure}{S\arabic{figure}}
\renewcommand{\theHfigure}{SM.\arabic{figure}}
\setcounter{table}{0}
\renewcommand{\thetable}{S\arabic{table}}
\renewcommand{\theHtable}{SM.\arabic{table}}
\section*{Supplemental Material for ``Chaos without Local Spectral Instability in a Kicked Oscillator''}

Equations are numbered independently of the Letter.
Equation, figure, and table numbers prefixed by ``S'' refer to this Supplement; unprefixed numbers refer to the Letter. As in the Letter, $\lambda=\e^{-\gamma T}$ is the damping factor per kick and $K$ the dimensionless kick strength; $\lambda_1\geq\lambda_2$ denote the Lyapunov exponents. The Supplement separates three kinds of results:
\begin{itemize}
\item \emph{exact analytic results}: basic properties, symmetries, and local spectral bounds of the return map (Sec.~S2), the analytic boundaries of Fig.~\ref{fig:phase}(a) (Sec.~S3), the exact joint spectral radius (Sec.~S4), finite-product identities (Sec.~S5), the Lyapunov-exponent decomposition (Sec.~S6), and the Lur'e formulation (Sec.~S9);
\item \emph{numerical diagnostics}: the parameter scan of Fig.~\ref{fig:phase}(b) (Sec.~S3), finite-product rates, pumping statistics, reordering experiments, basins of attraction, and the route to intermittency (Sec.~S7);
\item a \emph{computer-assisted proof}, by interval arithmetic, of a horseshoe with positive topological entropy (Sec.~S8).
\end{itemize}
Section~S1 summarizes the physical picture, Sec.~S10 proposes an experimental test, and Sec.~S11 separates the mechanical-energy balance from the tangent-space balance.

\section*{S1. GEOMETRIC AMPLIFICATION IN THE KICKED OSCILLATOR}
The mechanism identified in the Letter has a simple mechanical
interpretation. Each pulse from the periodic substrate shears
momentum, while the damped oscillation between successive pulses
rotates the resulting shear in phase space. Although every local
Jacobian has spectral radius $\sqrt{b}<1$, the product of
successive Jacobians need not contract, because the sign and strength of each shear, set by the local substrate curvature $t_n=\mu\cos y_n$, depend on the state reached at the corresponding kick.
When the trajectory visits restoring and anti-restoring regions
of the substrate ($t_n<0$ and $t_n>0$, respectively) in a suitable sequence, the corresponding tangent
dynamics can amplify perturbations exponentially. Eq.~\eqref{eq:two}
expresses this mechanism at the level of two successive kicks.

The joint spectral radius in Eq.~\eqref{eq:jsr} and the Lyapunov decomposition
in Eq.~\eqref{eq:lambda} describe different aspects of this amplification.
The former gives the largest asymptotic growth rate allowed by
arbitrary products from the Jacobian family. The alternating
sequence that attains this bound need not, however, be generated
by any trajectory of the oscillator; at the working point, the measured $\lambda_1\simeq0.103$ lies far below the bound $\ln q\simeq0.756$ (Sec.~S4). The Lyapunov decomposition,
in contrast, accounts for the sequence of Jacobians selected by
the actual nonlinear dynamics and quantifies the geometric
amplification accumulated along an orbit. The computer-assisted
covering construction provides a separate result: it proves the
existence of a compact invariant set with positive topological
entropy. This establishes chaotic invariant dynamics, but does
not by itself show that the numerically observed strange attractor
is attracting. Its attracting character is instead supported by
the numerical basin analysis and long-time persistence tests.

The same mechanism is reflected in the bifurcations of the return
map. The monodromy matrix of a period-$p$ orbit, a product of $p$ locally elliptic Jacobians with determinant $b^p$, can develop unstable
Floquet multipliers as the loading $C$ varies. The ensuing
sequence of collisions and bifurcations produces a saddle-node bifurcation of a period-24
orbit and the associated type-I intermittent dynamics.
Close to the bifurcation, the intermittent episodes reported in
Fig.~\ref{fig:route} are observed in chaotic transients. Sustained intermittent
chaos occurs in the parameter regime specified in Sec.~S7, $C-C_f\gtrsim8\times10^{-5}$.

The amplification of perturbations in tangent space should not
be confused with the mechanical energy supplied to the oscillator.
The optical or mechanical pulses perform work that compensates
for viscous dissipation, whereas the growth of infinitesimal
trajectory separations depends on how the successive substrate
curvatures orient the local contraction metrics. Accordingly,
the contribution $W_n$ in Eq.~\eqref{eq:lambda} is a term in the tangent-space
balance, not an amount of mechanical energy injected by a pulse.
Section~S11 quantifies the distinction between these two balances, using the explicit form of the contraction metric in the oscillator variables derived in Sec.~S6.

The same system can also be formulated as a Lur'e feedback
problem (Sec.~S9), providing a connection with questions of
absolute stability. Beyond laser-cooled trapped ions, possible
experimental settings include pulsed nanomechanical oscillators
and driven spring--block friction systems. Section~S10 describes
a possible experimental protocol and discusses the assumptions
and limitations of the impulsive, damped, and effectively
classical description.

\section*{S2. BASIC PROPERTIES AND SPECTRAL STABILITY OF THE RETURN MAP}

This section establishes the basic properties of the return map used in the Letter. We show that the map is invertible and that its forward orbits remain bounded, determine when the fixed point is unique and nontrivial period-two orbits are excluded, and derive conditions for spectral stability of the local Jacobians. These results provide the setting for the central question of the Letter: how products of individually stable Jacobians can nevertheless amplify perturbations.

For the quarter-period reduction, write $F(u,v)=(v,f(v)-bu+C)$ with $(u,v)=(y_{n-1},y_n)$, where $f(v)=\mu\sin v$ and $0<b<1$. The inverse map is $F^{-1}(v,w)=((f(v)+C-w)/b,v)$, so $F$ is a global real-analytic diffeomorphism. Moreover, since $|f(v)|\leq\mu$, the even ($r=0$) and odd ($r=1$) subsequences of iterates obey the bound
\begin{equation}\label{eq:SMbounded}
|y_{r+2j}|\leq b^j|y_r|+(\mu+|C|)\frac{1-b^j}{1-b},\qquad r\in\{0,1\},\quad j\geq0.
\end{equation}
Thus every forward orbit remains bounded and, for every $\varepsilon>0$, eventually enters the absorbing square $|u|,|v|\leq(\mu+|C|)/(1-b)+\varepsilon$.

The fixed point satisfies $(1+b)y_* - \mu\sin y_*=C$. For $\mu<1+b$, the derivative of the left-hand side is at least $1+b-\mu>0$, so the fixed-point equation has exactly one solution. The same condition rules out a nontrivial two-cycle. Indeed, a two-cycle with distinct values $u$ and $v$ would require $(1+b)(u-v)=-\mu(\sin u-\sin v)$, which is impossible for $u\neq v$ because $|\sin u-\sin v|\leq|u-v|$ and $\mu<1+b$. Both conclusions hold at the working point, where $\mu=1.7076<1+b=1.9$.

\emph{Symmetries.} The substitutions $y_n\to-y_n$ and $y_n\to y_n+2\pi$ map the recurrence~\eqref{eq:rec} with loading $C$ onto the same recurrence with loading $-C$ and $C+2\pi(1+b)$, respectively. The dynamics at loadings $C$, $-C$, and $C+2\pi(1+b)$ are therefore conjugate, and $0\leq C\leq\pi(1+b)$ is a fundamental domain for the loading, with $\pi(1+b)\simeq5.97$ at $b=0.9$.

The Jacobian $J(t)$ has characteristic polynomial
\begin{equation}\label{eq:SMschur}
\chi_t(z)=z^2-tz+b,\qquad \rho(J(t))<1\quad\text{for all }|t|\leq\mu\text{ if }\mu<1+b.
\end{equation}
The second statement is the Schur criterion, $b<1$ and $|t|<1+b$, for $\chi_t$. In particular, when $\mu<2\sqrt{b}$, its eigenvalues are complex conjugates with modulus $\sqrt{b}<1$ at every point in phase space.

The same local spectral stability can be ensured when the pulse spacing is detuned from a quarter of the damped period. For the Jacobian associated with Eq.~\eqref{eq:physical}, the determinant is $\lambda^2$ and the trace is $\lambda[2\cos\theta+K\cos(X+\varphi)\sin\theta]$. Its eigenvalues remain complex conjugates throughout phase space provided
\begin{equation}\label{eq:SMdetune}
K|\sin\theta|<2(1-|\cos\theta|),\qquad 0<\theta<\pi.
\end{equation}
Indeed, the eigenvalues are non-real if and only if $|2\cos\theta+Kc\sin\theta|<2$ for every $c=\cos(X+\varphi)\in[-1,1]$, that is, $2|\cos\theta|+K\sin\theta<2$. For $K<2$ this is equivalent to $K/2<\tan(\theta/2)<2/K$, the form quoted in the Letter; at $K=1.8$ it gives $83.97^\circ<\theta<96.03^\circ$. This bound gives the range of kicking phases and strengths for which every local Jacobian has eigenvalues of modulus $\lambda<1$ (here $\lambda=\e^{-\gamma T}$ is the damping factor, not a Lyapunov exponent), extending the spectral-stability condition beyond quarter-period driving. At fixed $\omega$ and $\gamma$, changing $T$ changes both $\theta=\omega T$ and $b=\lambda^2=\e^{-2\gamma T}$.

\section*{S3. ANALYTICAL BOUNDARIES AND NUMERICAL SCAN FOR FIG.~2}

Figure~\ref{fig:phase} separates two questions: which behaviours
are allowed by the family of local Jacobians, and which are
actually found along trajectories of the kicked oscillator.
Panel (a) gives the analytical boundaries, while panel (b) shows
the results of a numerical scan. The distinction matters because
the existence of amplifying Jacobian products does not by itself
imply that the nonlinear dynamics generates the sequences needed
to sustain such amplification.

In panel (a), the green region $\mu<1-b$ corresponds to
$q<1$ in Eq.~\eqref{eq:jsr}. In this range, every admissible
Jacobian product contracts in a common quadratic norm, and the
nonlinear map has a unique globally attracting fixed point. For
$1-b<\mu<2\sqrt b$, some products can instead amplify
perturbations even though every individual Jacobian has
complex-conjugate eigenvalues of modulus $\sqrt b<1$. This is
the parameter range relevant to the main result of the Letter.
At $\mu=2\sqrt b$, the eigenvalues can become real; nevertheless,
they remain inside the unit circle as long as $\mu<1+b$. The
dashed line at $\mu=1+b$ therefore marks the onset of possible
local spectral instability, not the onset of a real spectrum.
The boundaries in panel (a) follow from the analytical spectral
conditions rather than from numerical estimates.

Panel (b) fixes $b=0.9$. The green band, $0\leq\mu<0.1$,
indicates guaranteed global contraction. The vertical line at
$\mu=2\sqrt b\simeq1.897$ marks the end of the range with
complex eigenvalues at every point, while the local-instability
bound $\mu=1+b=1.9$ lies just beyond it. The horizontal line at
$\lambda_1=0$ separates regimes with negative and positive maximal
Lyapunov exponent. The negative plateau,
$\lambda_1=\tfrac12\ln b\simeq-0.0527$, corresponds to attracting
dynamics with complex Floquet multipliers: for a stable period-$p$ orbit (including $p=1$) whose multipliers are complex, $|\Lambda|=b^{p/2}$, so that $\lambda_1=\tfrac12\ln b$ exactly.

For each value of $\mu$, the black point in panel (b) gives the
largest value of $\lambda_1$ found over the sampled loading
values $C\in[0,6]$ and initial conditions in $[-8,8]^2$; by the symmetries of Sec.~S2, $C\in[0,6]$ covers all inequivalent loadings at $b=0.9$. The
coarse scan used a step of $0.062$ in $\mu$, 240 values of $C$,
and 16 initial conditions for each loading.
Exponents were calculated over $1.5\times10^4$ iterations following a transient
of the same length; runs of this length cannot distinguish an attractor from a long-lived chaotic transient (compare the chaotic saddle at $C=1.87$, with mean lifetime $\sim5\times10^6$ iterations, in Sec.~S7). Finer scans located a period-four attracting
orbit, created in a saddle-node bifurcation, near
$\mu\simeq0.36$, and chaotic attracting states with positive
maximal Lyapunov exponent near $\mu\simeq1.26$. The corresponding
windows in $C$ can be narrower than $0.02$. These values are the
onsets observed in the scans described here; they are not rigorous
thresholds for the existence of attracting or chaotic states.

The gap between the analytical boundary $\mu=1-b=0.1$ and the
numerically observed onset of chaos illustrates the difference
between possible amplification and amplification realised by the
nonlinear dynamics. Above $\mu=1-b$, the Jacobian family admits
expanding products, but the oscillator need not generate the
required ordering of substrate curvatures. Whether this potential
is realised depends on the sequences selected by the trajectory
and on the loading parameter. By contrast, the boundary
$\mu=2\sqrt b$ is an analytical limit for the local eigenvalue
spectrum, independent of the numerical scan and of the stability
of any particular attractor.

\section*{S4. EXACT JOINT SPECTRAL RADIUS AND ITS DYNAMICAL CONSEQUENCES}

The joint spectral radius measures the largest asymptotic growth
allowed by the full family of Jacobians, without requiring the
matrices to occur in an order generated by the oscillator. For
this family, the bound can be obtained exactly: a weighted norm
controls every finite product, and an alternating sequence of
extreme curvatures attains it. The same result bounds the Lyapunov
exponent along actual trajectories and identifies the parameter
range in which the return map is globally contracting.

For a bounded family of matrices $\mathcal J$, the joint spectral
radius is defined as
\begin{equation}\label{eq:SMjsrdef}
\widehat\rho(\mathcal J)=\lim_{L\to\infty}\sup_{A_j\in\mathcal J}
\norm{A_L\cdots A_1}^{1/L},
\end{equation}
where the limit exists and is independent of the chosen matrix norm. For
$\mathcal J=\mathcal J_{b,\mu}$${}=\{J(t):|t|\leq\mu\}$, introduce the similarity transform
$S=\mathrm{diag}(\sqrt b,1)$ and set
\begin{equation}\label{eq:SMB}
B(t)=SJ(t)S^{-1}=\begin{pmatrix}0&\sqrt b\\-\sqrt b&t\end{pmatrix},
\qquad \norm{J}_{b}=\norm{SJS^{-1}}_2.
\end{equation}
Here $\norm{\cdot}_b$ is the operator norm induced by the vector norm $\norm{(u,v)}_b=\norm{S(u,v)^{\mathsf T}}_2=(bu^2+v^2)^{1/2}$ of the Letter.
The singular values of $B(t)$ are
\begin{equation}\label{eq:SMsigma}
\sigma_\pm(B(t))=\frac{\sqrt{t^2+4b}\pm|t|}{2},
\qquad \sigma_+(B(t))\leq q(b,\mu).
\end{equation}
It follows that every product of $L$ admissible Jacobians has
weighted norm at most $q^L$, since $\sigma_+(B(t))$ increases with $|t|$ and the operator norm is submultiplicative. We now show that this bound is sharp
for every $L$, not just in the limit defining the joint spectral
radius.

Set $B_+=B(\mu)$ and $B_-=B(-\mu)$. These matrices satisfy
$B_-=-B_+^{\mathsf T}$. Writing the singular-value decomposition
as $B_+=U\Sigma V^{\mathsf T}$, we obtain
\[
(B_+B_-)^k=(-1)^kU\Sigma^{2k}U^{\mathsf T},
\qquad
(B_+B_-)^kB_+=(-1)^kU\Sigma^{2k+1}V^{\mathsf T}.
\]
The corresponding operator norms are $q^{2k}$ and $q^{2k+1}$,
respectively, because the largest entry of $\Sigma$ is $\sigma_+(B_+)=q$. Thus the alternating sequence of the two extreme
Jacobians attains the upper bound at every even and odd length,
so that
\begin{equation}\label{eq:SMfinite}
\sup_{|t_i|\leq\mu}\norm{J(t_L)\cdots J(t_1)}_b=q^L
\end{equation}
for each $L$. This proves the exact joint spectral radius stated
in Eq.~\eqref{eq:jsr}. In addition,
\begin{equation}\label{eq:SMpairattain}
\rho(J(\mu)J(-\mu))^{1/2}=q,\qquad
q<1\iff\mu<1-b.
\end{equation}
The spectral radius of the alternating two-factor product gives
a matching lower bound, while the weighted norm provides the
upper bound: $\rho(A_L\cdots A_1)^{1/L}\leq\widehat\rho(\mathcal J)\leq\sup\norm{A_L\cdots A_1}_b^{1/L}$ for every $L$ and all $A_j\in\mathcal J$.

The maximization above ranges over all admissible products. The
alternating sequence $t_i=\pm\mu$ that attains it need not be
realized by an orbit of the nonlinear map. Nevertheless, every Jacobian met along any orbit belongs to $\mathcal J_{b,\mu}$, so that $\norm{DF^n(z)}_b\leq q^n$ for all $z$ and $n$. Consequently, the
Lyapunov exponent associated with any invariant probability
measure satisfies
\begin{equation}\label{eq:SMlyapbound}
\lambda_1\leq\ln q(b,\mu).
\end{equation}
The distinction is important: the joint spectral radius describes
the largest growth allowed by the Jacobian family, whereas the
Lyapunov exponent measures the growth along the sequences generated
by the actual dynamics.

The same bound also gives a global contraction criterion. Let
$F_f(u,v)=(v,f(v)-bu+C)$, with $f\in C^1$ and
$\sup|f'|\leq\mu<1-b$. Because $F_f$ is nonlinear only through $f(v)$, one has exactly $F_f(z)-F_f(z')=J(\bar t)(z-z')$, with $\bar t=[f(v)-f(v')]/(v-v')\in[-\mu,\mu]$ by the scalar mean-value theorem ($\bar t$ may be chosen arbitrarily in $[-\mu,\mu]$ if $v=v'$). The case $L=1$ of Eq.~\eqref{eq:SMfinite} then gives
\begin{equation}\label{eq:SMcontract}
\norm{F_f(z)-F_f(z')}_b\leq q(b,\mu)\norm{z-z'}_b.
\end{equation}
Since $q<1$ in this range, Banach's fixed-point theorem guarantees
a unique globally attracting equilibrium. At the boundary
$\mu=1-b$, the map is nonexpansive in the same norm, and the
strict-contraction argument no longer settles the borderline case.

At the working parameters of the Letter,
$\ln q(0.9,1.7076)\simeq0.75617$, whereas the measured Lyapunov
exponent is $\lambda_1\simeq0.103$. The difference reflects the
fact that the nonlinear dynamics does not realize the most
amplifying sequences allowed by the full Jacobian family. Finally,
the interval $1-b<\mu<2\sqrt b$, in which expanding products are
possible while all local eigenvalues remain complex and inside the
unit circle, is nonempty precisely when $b>3-2\sqrt2$, since $1-b<2\sqrt b$ is equivalent to $(1+\sqrt b)^2>2$.

\section*{S5. FINITE-STEP PRODUCTS AND THE LIMITS OF THE TWO-KICK CRITERION}

The two-kick criterion in Eq.~\eqref{eq:two} identifies when a product of two successive Jacobians can expand spectrally. Longer products, however, can behave differently: their growth is not governed by the same requirement of alternating curvature signs. This section makes that distinction explicit.

The product of two Jacobians is
\begin{equation}\label{eq:SMtwo}
J(t_2)J(t_1)=\begin{pmatrix}-b&t_1\\-bt_2&t_1t_2-b\end{pmatrix},\qquad
\det=b^2,\quad\mathrm{tr}=t_1t_2-2b.
\end{equation}
For $0<b<1$ and $|t_i|<2\sqrt b$, spectral expansion requires
$|t_1t_2-2b|>1+b^2$: a real $2\times2$ matrix with determinant $b^2<1$ has an eigenvalue outside the unit circle if and only if its characteristic polynomial is negative at $z=1$ or at $z=-1$. The positive-trace branch would require
$t_1t_2>(1+b)^2$, which is incompatible with the global ellipticity
condition $|t_i|<2\sqrt b$, since then $t_1t_2<4b\leq(1+b)^2$. The remaining condition is therefore
$t_1t_2<-(1-b)^2$, as stated in Eq.~\eqref{eq:two}. Opposite-sign
curvatures are thus necessary for spectral expansion of a two-factor
product within this regime, but not for products of greater length.

For example, consider the admissible triple
\begin{equation}\label{eq:SMthree}
(t_1,t_2,t_3)=(0.1,1.7076,1.7076),\qquad b=0.9.
\end{equation}
Although neither of its consecutive two-factor products is spectrally
expanding,
$\rho(J(t_2)J(t_1))=\rho(J(t_3)J(t_2))=b<1$, the full product
satisfies $\rho(J(t_3)J(t_2)J(t_1))>2.5$: by Eq.~\eqref{eq:SMtr3}, its trace is $t_1t_2t_3-b(t_1+t_2+t_3)\simeq-2.872$ and its determinant is $b^3$, so that $\rho\simeq2.59$. All three curvatures are positive. This example shows that
longer products can amplify perturbations even when the two-step
criterion is never met and the curvature never changes sign. The triple is admissible for the Jacobian
family, but need not be generated by an actual trajectory.

The dependence on ordering becomes clearer at longer lengths. The
trace, and hence the spectrum, of a product of two or three such
companion matrices is unchanged by reordering its factors: for two factors by cyclicity of the trace, and for three because the trace, Eq.~\eqref{eq:SMtr3}, is a symmetric function of $t_1,t_2,t_3$, while the determinant equals $b^L$ in all cases. At
length four, this invariance no longer holds: changing the order
can change the spectral radius [Eqs.~\eqref{eq:SMtr4} and~\eqref{eq:SMswap}]. Along an actual orbit, the nonlinear
dynamics selects a particular sequence of curvatures, and this
ordering governs the growth of long Jacobian products and the
resulting Lyapunov exponent.

\section*{S6. EXACT GEOMETRIC-PUMPING BUDGET AND THE LYAPUNOV EXPONENT}

This section derives an exact balance for the growth of infinitesimal
perturbations along an orbit. Each local Jacobian contracts by a factor
$b_n$ in its associated quadratic metric, but that metric changes from
one step to the next. The resulting mismatch contributes to the net
growth of perturbations and gives an exact decomposition of the maximal
Lyapunov exponent into a local contraction term and a geometric term.
The construction also extends beyond the sinusoidal kick map to a
broader class of second-order recurrences.

For $b>0$ and $t^2<4b$, define the positive-definite quadratic metric
\begin{equation}\label{eq:SMG}
G(b,t)=\begin{pmatrix}b&-t/2\\-t/2&1\end{pmatrix},\qquad
\det G=b-\frac{t^2}{4}>0.
\end{equation}
Direct multiplication gives
$J(b,t)^\mathsf T G(b,t)J(b,t)=bG(b,t)$, as in
Eq.~\eqref{eq:metric}. Thus, in the metric associated with a given
Jacobian, the corresponding tangent vector contracts exactly by the
factor $b$ at the level of its squared norm.
Up to a positive factor, $G(b,t)$ is the only symmetric matrix satisfying this relation when $t^2<4b$: $J(b,t)/\sqrt b$ is then conjugate to a rotation by an angle in $(0,\pi)$, and the only quadratic forms invariant under such a rotation are multiples of the Euclidean one. The decomposition derived below is therefore canonical, up to the rescaling discussed at the end of this section.

To see how changes in this metric affect long-time growth, consider the
more general recurrence
\begin{equation}\label{eq:SMmemory}
y_{n+1}=f(y_n)-h(y_{n-1})+C,\qquad
 t_n=f'(y_n),\quad b_n=h'(y_{n-1}),
\end{equation}
where $0<b_-\leq b_n\leq b_+<1$ and
$|t_n|\leq\mu_0<2\sqrt{b_-}$ along the orbit. Every local Jacobian
therefore has complex-conjugate eigenvalues of modulus
$\sqrt{b_n}<1$, and the metrics $G_n=G(b_n,t_n)$ remain uniformly
positive definite and bounded. Let
$\xi_n=(\delta u_n,\delta v_n)^\mathsf T\neq0$ [written $(u_n,v_n)^{\mathsf T}$ in the Letter] satisfy $\xi_{n+1}=J_n\xi_n$, and define
$E_n=\xi_n^\mathsf T G_n\xi_n$. Inserting and subtracting
$\xi_{n+1}^\mathsf T G_n\xi_{n+1}$ yields the exact balance
\begin{align}\label{eq:SMbalance}
E_{n+1}&=\xi_{n+1}^\mathsf T G_n\xi_{n+1}
+\xi_{n+1}^\mathsf T(G_{n+1}-G_n)\xi_{n+1}\nonumber\\
&=b_n E_n+W_n,\\
W_n&=(b_{n+1}-b_n)(\delta u_{n+1})^2
-(t_{n+1}-t_n)\delta u_{n+1}\delta v_{n+1}.
\nonumber
\end{align}
The term $b_nE_n$ describes the contraction associated with the current
Jacobian, while $W_n$ accounts for the change in metric between
successive steps.

Since
$E_{n+1}/(b_nE_n)=1+W_n/(b_nE_n)>0$ (both quadratic forms are positive because the $G_n$ are positive definite and $J_n$ is invertible), taking logarithms and summing
along the orbit gives
\begin{equation}\label{eq:SMtelescoping}
\frac{1}{2N}\ln\frac{E_N}{E_0}
=\frac1{2N}\sum_{n=0}^{N-1}\ln b_n
+\frac1{2N}\sum_{n=0}^{N-1}
\ln\left(1+\frac{W_n}{b_nE_n}\right).
\end{equation}
Because the metrics are uniformly equivalent to the Euclidean metric,
there are constants  $0<c_-\leq c_+$ such that
$c_-\norm{\xi}^2\leq\xi^\mathsf T G_n\xi\leq c_+\norm{\xi}^2$
for all $n$ and all $\xi$. They are the smallest and largest
eigenvalues of $G(b,t)$ over the admissible range
$b_-\leq b\leq b_+$, $|t|\leq\mu_0$, namely
$c_\mp=\tfrac12\big[1+b_\mp\mp\sqrt{(1-b_\mp)^2+\mu_0^2}\big]$;
positivity of $c_-$ is equivalent to $\mu_0<2\sqrt{b_-}$.
Hence the difference between the left-hand side of
Eq.~\eqref{eq:SMtelescoping} and
$N^{-1}\ln(\norm{\xi_N}/\norm{\xi_0})$ is bounded in modulus by $(2N)^{-1}\ln(c_+/c_-)$ and vanishes as
$N\to\infty$. For an ergodic invariant measure $\nu$ and $\nu$-almost every initial state, the Oseledets theorem gives $N^{-1}\ln(\norm{\xi_N}/\norm{\xi_0})\to\lambda_1$ for every $\xi_0$ outside the slow Oseledets subspace (in particular for Lebesgue-almost every $\xi_0$), while the Birkhoff theorem gives convergence of the first average on the right-hand side of Eq.~\eqref{eq:SMtelescoping}, because $\ln b_n=\ln h'(y_{n-1})$ is a bounded continuous function of the state. The second average therefore converges as well, as the difference of two convergent sequences, and no further assumption is needed:
\begin{equation}\label{eq:SMgenlambda}
\lambda_1=\frac12\lim_{N\to\infty}\frac1N\sum_{n=0}^{N-1}\ln b_n
+\frac12\lim_{N\to\infty}\frac1N\sum_{n=0}^{N-1}
\ln\left(1+\frac{W_n}{b_nE_n}\right).
\end{equation}
This identity separates the average local contraction from the
contribution associated with the changing tangent metric.

For constant $b_n=b$, the first term reduces to
$\frac12\ln b$, while
$W_n=-(t_{n+1}-t_n)\delta u_{n+1}\delta v_{n+1}$.
Equation~\eqref{eq:SMgenlambda} then gives the decomposition in
Eq.~\eqref{eq:lambda}. Independently, the determinant of the Jacobian
product fixes the sum of the two Lyapunov exponents:
\begin{equation}\label{eq:SMsumlyap}
\lambda_1+\lambda_2=\langle\ln b_n\rangle,
\qquad\text{in particular }\lambda_1+\lambda_2=\ln b.
\end{equation}

For numerical evaluation, the logarithmic increments can be computed
without carrying the norm of the tangent vector by using
$\widehat\xi_n=\xi_n/\norm{\xi_n}$. In this form,
\begin{equation}\label{eq:SMnumerical}
\frac{W_n}{b_nE_n}=
\frac{(J_n\widehat\xi_n)^\mathsf T(G_{n+1}-G_n)(J_n\widehat\xi_n)}
{b_n\widehat\xi_n^\mathsf T G_n\widehat\xi_n}.
\end{equation}
The contribution depends on both the local curvature $t_n$ and the
orientation of the perturbation. It measures the mismatch between
successive tangent metrics, not the mechanical energy delivered by an
external pulse; the latter is treated separately in Sec.~S11. Finally,
rescaling the metric by a bounded positive factor changes the finite-step
balance only through terms that telescope, leaving the Lyapunov exponent
unchanged.

\emph{Physical form of the metric.} For the sinusoidally kicked oscillator at $\theta=\pi/2$, the map~\eqref{eq:physical} gives $P_n=-\lambda X_{n-1}$, so that $\xi_n=(\delta y_{n-1},\delta y_n)^{\mathsf T}=(-\delta P_n/\lambda,\,\delta X_n)^{\mathsf T}$, where $(\delta X_n,\delta P_n)$ is the perturbation in the rescaled phase-space variables of the Letter. With $t_n=\lambda K\cos y_n$, the quadratic form and the pumping term become
\begin{equation}\label{eq:SMphys}
E_n=\delta X_n^2+\delta P_n^2+K\cos y_n\,\delta X_n\,\delta P_n,\qquad
W_n=K\,(\cos y_{n+1}-\cos y_n)\,\delta X_{n+1}\,\delta P_{n+1}.
\end{equation}
The first two terms of $E_n$ form the squared amplitude of the free damped oscillation, which the damped rotation $\lambda R(\pi/2)$ reduces exactly by the factor $b$ between kicks; the cross term is the tilt imposed by the kick shear $K\cos y_n$. Geometric pumping is thus the product of the change of local shear between successive kicks and the position--momentum correlation $\delta X_{n+1}\delta P_{n+1}$ of the perturbation.

\emph{Slowly varying curvature.} For constant $b$, the balance~\eqref{eq:SMbalance} involves only the matrices $J(t_n)$ and therefore holds for any sequence of admissible Jacobians, whether or not it is generated by an orbit. Since $bE_n=\xi_{n+1}^{\mathsf T}G_n\xi_{n+1}$, $|\delta u_{n+1}\delta v_{n+1}|\leq\norm{\xi_{n+1}}^2/2$, and $\xi^{\mathsf T}G_n\xi\geq g_{\min}\norm{\xi}^2$, where $g_{\min}=\tfrac12[1+b-\sqrt{(1-b)^2+\mu^2}]$ is the smallest eigenvalue of $G(t)$ for $|t|\leq\mu$ (the constant $c_-$ above), each increment satisfies
\begin{equation}\label{eq:SMslow}
\ln\left(1+\frac{W_n}{bE_n}\right)\leq\frac{|t_{n+1}-t_n|}{2g_{\min}}.
\end{equation}
Geometric pumping therefore requires curvature changes: if $\sum_{n<N}|t_{n+1}-t_n|$ grows sublinearly in $N$, the growth rate of the product reduces to $\tfrac12\ln b$, which is also its lower bound set by the determinant. Section~S7 uses this bound to explain the reordering experiment of the Letter.

\section*{S7. ORDERED SPECTRAL GROWTH AND NUMERICAL DIAGNOSTICS}

We use several complementary diagnostics to quantify how the ordering of
successive Jacobians affects perturbation growth. The first measures
the spectral radius of finite products along an orbit; the others
compare the geometric-pumping budget with direct Lyapunov estimates, reorder the Jacobians sampled along the attractor,
and examine which curvature sequences contribute most to the growth.
We also report the numerical tests used to establish the robustness
of the chaotic attractor and to characterize the route to
intermittency.

The finite-product diagnostic introduced in
Ref.~\cite{Sornette2026arXiv} is
\begin{equation}\label{eq:SMhL}
h_L=\frac1L\left\langle\ln\rho(J_{n+L-1}\cdots J_n)\right\rangle.
\end{equation}
At a fixed point with an elliptic Jacobian, this quantity is
$h_L=\ln\rho(J)=\tfrac12\ln b$ for every $L$. Along the chaotic
orbit considered in the Letter, the one-step value is the same,
$h_1=\tfrac12\ln b$, but the two-step value is positive,
$h_2\simeq0.255$. For the range of longer products examined here, $h_L$ approaches
the directly computed top Lyapunov exponent.

\begin{figure}[h]
\centering
\includegraphics[width=0.86\textwidth]{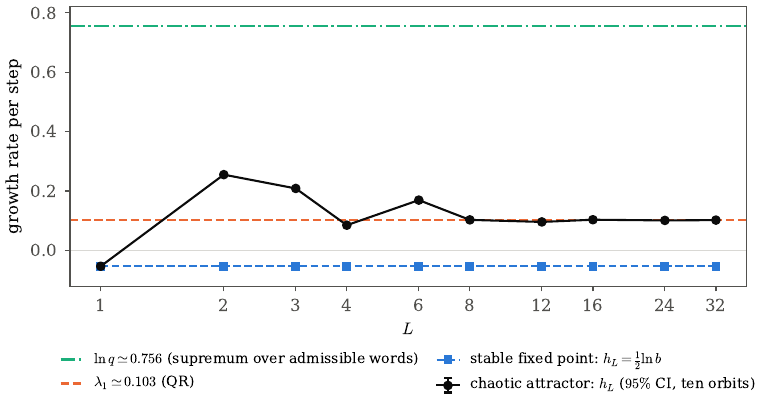}
\caption{Finite-product rate $h_L$ for the chaotic orbit, compared with the fixed-point value, the measured maximal Lyapunov exponent, and the exact upper growth rate $\ln q$ for the full admissible Jacobian family. Every local Jacobian has spectral radius $\sqrt b$, but the supremum over all admissible products need not be attained along the dynamics.\label{fig:SMhL}}
\end{figure}

The effect of ordering can be seen directly from the trace of the
matrix products (see also Sec.~S5). For arbitrary $t_1,t_2,t_3$, one finds
\begin{equation}\label{eq:SMtr3}
\operatorname{tr}[J(t_3)J(t_2)J(t_1)]
=t_1t_2t_3-b(t_1+t_2+t_3),
\end{equation}
which is invariant under permutations of the three factors. At
four steps, however,
\begin{align}\label{eq:SMtr4}
\operatorname{tr}[J(t_4)J(t_3)J(t_2)J(t_1)]
&=t_1t_2t_3t_4+2b^2\nonumber\\
&\quad-b(t_1t_2+t_2t_3+t_3t_4+t_4t_1),
\end{align}
and swapping the adjacent factors associated with $t_2$ and $t_3$
changes the trace by
\begin{equation}\label{eq:SMswap}
\Delta\operatorname{tr}=b(t_1-t_4)(t_2-t_3),
\end{equation}
where $\Delta\operatorname{tr}$ denotes the new trace minus the
original one. Since every four-factor product has determinant
$b^4$, its spectral radius can therefore depend on the order of
the factors. The orbit fixes one particular ordering among all
the products allowed by the Jacobian family.

Sorting and shuffling the Jacobians provide a way to isolate this
effect numerically. Both procedures preserve the multiset of
matrices, and hence all one-step statistics, while changing the
order in which the matrices are multiplied. The resulting products contain the same local Jacobians, but in a different order from the one generated by the nonlinear trajectory. The corresponding growth rates are discussed in item~(v) below.

The diagnostics below refer to the attractor at
$(b,\mu,C)=(0.9,1.7076,1.885)$. They were computed from ten
independent orbits of $10^6$ iterations, each preceded by a
transient of $10^5$ iterations. Unless stated otherwise, the
uncertainties are $95\%$ confidence intervals across the ten orbits.

\emph{(i) Pumping contribution.}
The mean pumping contribution to the Lyapunov budget is
$0.1557\pm0.0002$ across the ten orbits, and
$0.1555\pm0.0007$ when estimated from $32$ contiguous blocks of
a single orbit. Adding $\tfrac12\ln b$ gives $0.1030$. This
agrees with the growth rate of the same tangent vector to within
$1.3\times10^{-6}$, the bound on the finite-time boundary term
in the telescoping sum,
$(2N)^{-1}\ln\operatorname{cond}G=1.5\times10^{-6}$, where $\operatorname{cond}G=c_+/c_-\simeq19.06$ is the largest condition number of $G(t)$ for $|t|\leq\mu$ (Sec.~S6), and with
the QR estimate $\lambda_1=0.1030\pm0.0002$ to within
$6\times10^{-5}$ on every orbit. The pumping increments were
recorded after a burn-in of $10^4$ iterations, allowing the tangent
vector to align with the top Oseledets direction. The two averages
therefore use slightly different time windows, and their small
difference is a finite-time fluctuation. The QR exponents also
satisfy $\lambda_1+\lambda_2=\ln b$ to $10^{-14}$.

\emph{(ii) Dependence on successive curvatures.}
Table~\ref{tab:SMpump} groups the pumping increments according to
the sign and product of successive curvatures, $t_nt_{n+1}$.
Pairs that satisfy the two-step criterion in Eq.~\eqref{eq:two}
account for $49.8\%$ of all steps but contribute $83\%$ of the
total pumping. Their mean increment is $0.261$, and the increment
is positive in $71\%$ of these cases. Pairs with curvatures of the
same sign account for the remaining $17\%$ of the pumping.
Opposite-sign pairs that do not meet the two-step criterion are
rare and make a negligible contribution. Thus, the pairs that
expand spectrally are also the dominant source of geometric
pumping.

\emph{(iii) Finite-length spectral rates.}
Table~\ref{tab:SMhL} gives $h_L$ and its uncertainty, estimated
both across the ten independent orbits and across $32$ contiguous
blocks of one orbit. The uncertainties are much smaller than the
differences between successive values of $L$. The non-monotonic
dependence on product length is therefore a reproducible feature
of the ordered products, rather than a fluctuation within the
estimated uncertainties.

\begin{figure}[h]
\centering
\includegraphics[width=0.95\textwidth]{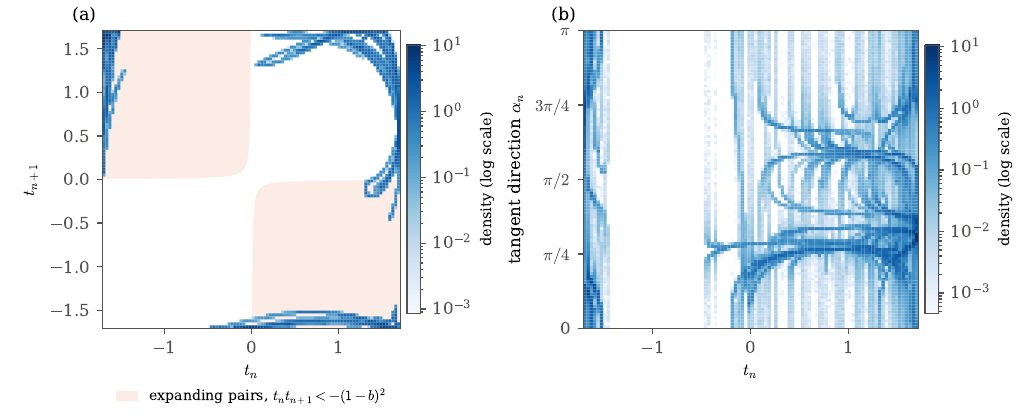}
\caption{Empirical joint densities on the attractor at $C=1.885$, computed from four orbits of $5\times10^5$ iterations and shown on a logarithmic scale. (a) Successive curvatures $(t_n,t_{n+1})$. The shaded regions, bounded by $t_nt_{n+1}=-(1-b)^2$, contain the expanding pairs satisfying Eq.~\eqref{eq:two}. (b) Curvature $t_n$ and orientation $\alpha_n\in[0,\pi)$ of the top Oseledets vector.\label{fig:SMjoint}}
\end{figure}

\emph{(iv) Joint distributions.}
Figure~\ref{fig:SMjoint} shows the empirical joint density of
successive curvatures $(t_n,t_{n+1})$ and that of the curvature
$t_n$ with the orientation $\alpha_n$ of the top Oseledets vector.
Successive curvatures cluster near $t=\pm\mu$ and occur frequently
in the second and fourth quadrants, where Eq.~\eqref{eq:two}
holds. The orientation of the tangent direction is also strongly
correlated with the curvature. The Lyapunov exponent therefore
depends on the joint evolution of the curvature sequence and the
direction of the perturbation.

\emph{(v) Reordering experiment.} The Letter multiplies the same $N=2\times10^5$ Jacobians, sampled along the attractor, in three orders and reports the growth rate $N^{-1}\ln\norm{J_{N-1}\cdots J_0}$ (spectral norm): $0.1027\pm0.0006$ in chronological order, which reproduces $\lambda_1$; $0.1857\pm0.0010$ after a random permutation; and $-0.0527$ after sorting by $t_n$.
The sorted value follows from Eq.~\eqref{eq:SMslow}. For a monotone sequence, $\sum_n|t_{n+1}-t_n|\leq2\mu$, so the pumping term contributes at most $\mu/(2g_{\min}N)$ to the growth rate. Together with the boundary term of Sec.~S6 and the determinant bound $\norm{J_{N-1}\cdots J_0}\geq b^{N/2}$, this confines the growth rate of the sorted product to the interval $[\tfrac12\ln b,\,\tfrac12\ln b+\mu/(2g_{\min}N)+(2N)^{-1}\ln\operatorname{cond}G]$, of width $5\times10^{-5}$ for $N=2\times10^5$. Sorting thus suppresses geometric pumping altogether, although every one-step property of the Jacobians is unchanged.

\emph{Robustness and choice of the working point.}
At $C=1.885$, all $64$ trajectories followed for $3\times10^7$
iterations remain chaotic. Their finite-time Lyapunov exponents,
computed over windows of $10^5$ iterations, stay between $0.0995$
and $0.106$. This persistence makes $C=1.885$ a suitable working
point. Elsewhere in the chaotic window, the chaotic set can be
transient. At $C=1.87$, for example, it is a long-lived chaotic
saddle: $163$ of $256$ trajectories leave it within $5\times10^6$
iterations, and all escaping trajectories converge to the stable
fixed point. The estimated mean lifetime is about $5\times10^6$
iterations, the maximum-likelihood estimate obtained with an
exponential escape law and right-censoring at $5\times10^6$
iterations.

To estimate the basins at the working point, an initial condition
is classified as converging to the fixed point if its final
distance from $y_*$ is below $10^{-6}$, and as belonging to the
chaotic attractor if its finite-time exponent over $2\times10^4$
iterations exceeds $0.02$. The remaining initial conditions
converge to periodic orbits of periods $8$, $16$, or $24$. The
basin fractions were measured on grids covering $[-4,5]^2$.
Because the basins are interleaved, shifting this square by one
unit changes the estimated chaotic fraction by up to three
percentage points. On the $81\times81$ grid, the fractions are
$(44.6\pm0.6)\%$, $(36.9\pm0.6)\%$, and $(18.4\pm0.5)\%$ for
the fixed point, chaotic attractor, and other periodic attractors,
respectively. These values are unchanged when the transient is
increased from $2\times10^4$ to $4\times10^4$ iterations. On a
$161\times161$ grid, the corresponding fractions are
$(44.4\pm0.3)\%$, $(37.2\pm0.3)\%$, and $(18.5\pm0.2)\%$.
The quoted uncertainties in these basin fractions are binomial
standard errors; they are much smaller than the systematic dependence on the sampling square noted above, so the fractions characterize the chosen square rather than intrinsic basin measures.

\emph{Numerics of the route to intermittency.}
Periodic orbits of period $p$ were found by applying Newton's
method to the closed $p$-periodic recurrence. Iterations were
stopped when the correction fell below $10^{-13}$ or after fifty
steps, and solutions were retained only when their residuals were
below $10^{-12}$. The period-12 branch was continued in $C$ with
steps of $1.8\times10^{-6}$. The collision of its Floquet
multipliers was bracketed by a sign change in the monodromy
discriminant $(\operatorname{tr}M)^2-4\det M$ and refined by bisection to $10^{-14}$, giving
$C^*=1.8428474758$. The period-doubling point was located by
bisecting $\det(M+I)$ to $10^{-15}$, yielding
$C=1.8448821718$.

The period-24 branch was then continued with an adaptively halved
step, down to $10^{-13}$. The last converged point and the
extrapolation of $(1-\Lambda)^2$ to zero agree to
$5\times10^{-12}$, giving $C_f=1.8459300302$. The scaling
$1-\Lambda\propto(C_f-C)^{0.497}$ over
$10^{-10}<C_f-C<10^{-5}$ confirms the saddle-node; as the fitting
window approaches $C_f$, the fitted exponent approaches $1/2$.

To characterize the intermittency, $16$ independent trajectories
were followed for $2\times10^7$ iterations at each of nine values
of $C-C_f$ between $10^{-8}$ and $10^{-4}$. A laminar phase is
defined as a stretch longer than three periods of the period-24 orbit (72 iterations) for which
$|y_n-y_{n-24}|<\delta$.
Each trajectory starts within $10^{-3}$
of the period-24 orbit at $C_f$. For $C-C_f<7.6\times10^{-5}$,
the intermittent dynamics is transient, because it coexists with stable periodic orbits. A stable period-28 orbit
is born at a saddle-node bifurcation at
$C_f+7.6\times10^{-5}$ and undergoes a period-doubling cascade
as $C$ decreases, accumulating at $C_f-3.2\times10^{-6}$. Each
trajectory is consequently captured by an orbit of period $28$,
$56$, or $112$ after $10^4$ to $10^6$ iterations, after which
its Lyapunov exponent is negative. These periodic orbits do not
meet the laminar-phase criterion, so the laminar phases are
collected from the chaotic transient, with between $450$ and
$800$ phases per value of $C$.

Sustained intermittent chaos, with positive Lyapunov exponent,
is found for $C-C_f\gtrsim8\times10^{-5}$. For $\delta=0.02$, a
weighted fit over $C-C_f\leq3.2\times10^{-5}$ gives the scaling
exponent $-0.500\pm0.002$, where the weights are the inverse
variance of the logarithm of the mean laminar time. The results
are robust to the choice of threshold: for $\delta=0.01$ and
$0.04$, the fitted exponents are $-0.513\pm0.004$ and
$-0.496\pm0.002$, respectively. The largest distance,
$C-C_f=10^{-4}$, is excluded from the fit because the mean
laminar time is then only about $400$ iterations, too short to
separate laminar phases reliably from chaotic bursts.\par

\begin{table}[h]
\centering
\caption{Pumping increments $\tfrac12\ln(1+W_n/bE_n)$ classified by the product of successive curvatures, for ten orbits of $10^6$ iterations at $C=1.885$.\label{tab:SMpump}}
\begin{tabular}{lcccc}
\hline
Class & Fraction (\%) & Mean increment & Positive (\%) & Share of total (\%)\\
\hline
Expanding pair, $t_nt_{n+1}<-(1-b)^2$ & 49.8 & $+0.261$ & 71 & $+83.3$ \\
Opposite sign, non-expanding & 0.2 & $-0.056$ & 51 & $-0.1$ \\
Same sign, $t_nt_{n+1}\geq0$ & 50.0 & $+0.052$ & 51 & $+16.7$ \\
\hline
\end{tabular}
\end{table}

\begin{table}[h]
\centering
\caption{Ordered-product rate $h_L$ at $C=1.885$: $95\%$ confidence intervals across ten independent orbits ($3.2\times10^5$ products with consecutive starting points each), and across $32$ contiguous blocks of a single orbit. The QR exponent is $\lambda_1=0.1030\pm0.0002$.\label{tab:SMhL}}
\begin{tabular}{ccc}
\hline
$L$ & $h_L$ (ten orbits) & $h_L$ (32 blocks)\\
\hline
1 & $-0.0527$ & $-0.0527$ \\
2 & $0.2554\pm0.0001$ & $0.2554\pm0.0002$ \\
3 & $0.2090\pm0.0003$ & $0.2092\pm0.0007$ \\
4 & $0.0854\pm0.0002$ & $0.0852\pm0.0005$ \\
6 & $0.1697\pm0.0002$ & $0.1697\pm0.0004$ \\
8 & $0.1033\pm0.0002$ & $0.1033\pm0.0006$ \\
12 & $0.0964\pm0.0003$ & $0.0964\pm0.0007$ \\
16 & $0.1038\pm0.0002$ & $0.1037\pm0.0006$ \\
24 & $0.1015\pm0.0002$ & $0.1013\pm0.0008$ \\
32 & $0.1025\pm0.0002$ & $0.1022\pm0.0007$ \\
\hline
\end{tabular}
\end{table}

\section*{S8. COMPUTER-ASSISTED CHAOS AND POSITIVE LYAPUNOV EXPONENTS}

The covering argument establishes the existence of chaotic invariant
 dynamics for the nonlinear map. It uses $h$-sets, parallelograms with one expanding (exit) and one contracting (entry) direction in the sense of Ref.~\cite{ZgliczynskiGidea2004} (see the Certificate paragraph below), around a period-four
saddle and along a near-homoclinic excursion. The 63 covering relations
verified by interval arithmetic form two return loops, of lengths 4 and
60. Repeating the shorter loop 15 times produces a second return path of
length 60, equal to that of the longer loop. Since these two paths can be concatenated in any order, the
construction allows two choices every 60 iterations and gives a symbolic
description of the dynamics on a compact invariant set.

More precisely, the covering construction yields a compact invariant
set $\mathcal K$ for $F^{60}$ and a continuous surjective coding
$\pi:\mathcal K\to\Sigma_2$, where $\Sigma_2$ is the space of bi-infinite
sequences of two symbols. The coding satisfies
\begin{equation}
\pi\circ F^{60}=\sigma\circ\pi,
\label{eq:SMcoding}
\end{equation}
with $\sigma$ the shift map. Each block of 60 iterations therefore
encodes one of two alternatives, and every bi-infinite sequence of these
alternatives is represented by an orbit in $\mathcal K$. After $m$ blocks, there
are $2^m$ distinct symbolic histories. It follows that
\begin{equation}
h_{\mathrm{top}}(F^{60}|_{\mathcal K})\geq\ln 2,
\label{eq:SMentropy56}
\end{equation}
because topological entropy cannot increase under a factor map. For the compact $F$-invariant set $\widetilde{\mathcal K}=\bigcup_{j=0}^{59}F^j(\mathcal K)$, this gives $h_{\mathrm{top}}(F|_{\widetilde{\mathcal K}})=\tfrac1{60}h_{\mathrm{top}}(F^{60}|_{\widetilde{\mathcal K}})\geq\ln2/60$, as stated in
Eq.~\eqref{eq:entropy} of the Letter.

The same construction yields a quantitative bound on the Lyapunov
exponent of an invariant measure. Consider the Bernoulli measure on
$\Sigma_2$ that assigns equal probability to the two symbols. Its metric
entropy per shift is $\ln 2$. Since the symbolic dynamics is a factor of
$F^{60}|_{\mathcal K}$, this measure can be lifted (a $\sigma$-invariant measure always admits an $F^{60}$-invariant preimage under the continuous surjection $\pi$, and the entropy of an extension is at least that of its factor) to an $F^{60}$-invariant
probability measure $\nu$ on $\mathcal K$ whose entropy is at least as large:
\begin{equation}
h_\nu(F^{60})\geq\ln 2.
\label{eq:SMlift56}
\end{equation}

To obtain an invariant measure for the original map $F$, average the
images of $\nu$ over the 60 intermediate iterates,
\begin{equation}
\overline{\nu}
=\frac{1}{60}\sum_{j=0}^{59}F_*^j\nu,
\label{eq:SMaverage}
\end{equation}
where $F_*^j\nu$ denotes the measure obtained by evolving $\nu$ for
$j$ iterations. Because $\nu$ is invariant under $F^{60}$, the average
$\overline{\nu}$ is invariant under $F$. Abramov's formula, $h_{\overline\nu}(F^{60})=60\,h_{\overline\nu}(F)$, together with $h_{\overline\nu}(F^{60})=h_\nu(F^{60})$ (entropy is affine in the measure, and each $F^j_*\nu$ is isomorphic to $\nu$ through $F^j$), then gives
\begin{equation}
h_{\overline{\nu}}(F)
=\frac{1}{60}h_\nu(F^{60})
\geq\frac{\ln 2}{60}.
\label{eq:SMlift}
\end{equation}

We next use this entropy bound to infer exponential growth of tangent
perturbations. By the ergodic decomposition theorem, at least one
ergodic component $\eta$ of $\overline{\nu}$ satisfies
$h_\eta(F)\geq\ln 2/60$. Ruelle's inequality~\cite{Ruelle1978} bounds metric entropy by
the sum of positive Lyapunov exponents. Since this is a two-dimensional
map and $\det DF=b<1$, its Lyapunov exponents obey
$\lambda_1+\lambda_2=\ln b<0$. Hence at most one exponent can be
positive, and Ruelle's inequality gives
\begin{equation}
\lambda_1(\eta)
\geq h_\eta(F)
\geq\frac{\ln 2}{60}
\simeq 0.0116>0.
\label{eq:SMpos}
\end{equation}
The covering proof thus guarantees an ergodic invariant measure with a
strictly positive Lyapunov exponent and a quantitative lower bound on
its asymptotic tangent-space growth. This result complements the
numerically measured value $\lambda_1\simeq0.103$ on the observed
chaotic attractor; the proof establishes the existence of a positive-entropy invariant
set and does not, by itself, identify its measure
with the physical measure of that attractor.

\paragraph{Certificate.}
The verification uses parallelogram $h$-sets of the form
$N=c+M[-1,1]^2$, with the centre $c$ and matrix $M$ specified exactly
as binary floating-point numbers. The first coordinate is expanding,
with exit edges $u=\pm1$, while the second is contracting, with entry
edges $s=\pm1$. For a proposed relation $N\Rightarrow N'$, define
\[
(u',s')=M'^{-1}[F(c+M(u,s))-c'],
\]
where $M'^{-1}$ is enclosed rigorously in interval arithmetic.
We verify two sufficient conditions: (A) $|s'|<1$ throughout the image
of $N$; and (B) the images of the exit edges $u=-1$ and $u=+1$ lie in
opposite half-planes, $\{\epsilon_N u'<-1\}$ and $\{\epsilon_N u'>1\}$ respectively, with $\epsilon_N=+1$ for an orientation-preserving and $\epsilon_N=-1$ for an orientation-reversing relation. Under these conditions,
the straight-line homotopy from $(u',s')$ to $(2\epsilon_N u,0)$ does not meet
the entry set of $N'$ and keeps the exit set outside $N'$: along the homotopy, the second coordinate stays in $(-1,1)$ by (A), and on the exit edges the first coordinate stays beyond $\pm1$ by (B). This is the
definition of a covering relation with one expanding direction
\cite{ZgliczynskiGidea2004}.

Both conditions are checked using multiple-precision interval arithmetic
with an 80-bit mantissa. The decimal parameters $b=0.9$, $\mu=1.7076$,
and $C=1.885$ are enclosed in rigorous intervals. Each $h$-set is
subdivided into $64\times16$ boxes, and each exit edge into 256 segments.
Whenever a condition cannot be decided on a box or segment, it is
bisected, up to eight times. The complete verification requires about
$1.3\times10^7$ interval evaluations.

The chain contains four $h$-sets, $P_0,\ldots,P_3$, around the period-four
saddle, whose multipliers are $-2.306$ and $-0.285$, together with 58
sets $X_9,\ldots,X_{66}$ along the near-homoclinic excursion. The 63
covering relations form two loops that share the final link
$P_3\Rightarrow P_0$:
\[
P_0\Rightarrow P_1\Rightarrow P_2\Rightarrow P_3\Rightarrow P_0
\quad\text{(loop A, length 4)},
\]
and
\[
P_0\Rightarrow X_9\Rightarrow\cdots\Rightarrow X_{66}
\Rightarrow P_3\Rightarrow P_0
\quad\text{(loop B, length 60)}.
\]
The two symbolic words are $A^{15}$ and $B$, both of length 60. At the
40th iterate, the path $A^{15}$ lies in $N(P_0)$, whereas $B$ lies in
$N(X_{48})$. A rigorous separating-axis test confirms that these two
parallelograms are disjoint, ensuring that the coding $\pi$ is
well-defined and continuous. Explicitly, the $j$th symbol of $\pi(x)$ is $A^{15}$ or $B$ according to whether $F^{60j+40}(x)$ lies in $N(P_0)$ or in $N(X_{48})$; since these are disjoint compact sets, the assignment is unambiguous and depends continuously on $x$.

For every covering relation, the margins $1-\sup|s'|$ and
$\inf|u'|-1$ in conditions (A) and (B) are strictly positive. The
verification records the slack in the interval enclosures at the final
subdivision, providing certified lower bounds on these margins. The
smallest such bounds are $2.7\times10^{-8}$ and $5.7\times10^{-6}$,
respectively. These small values measure how narrowly the interval enclosure of the most difficult box or edge segment passes the test, not how close the dynamics comes to violating it: the minimum margins themselves, enclosed separately, are $7.4\times10^{-2}$ and $3.1\times10^{-2}$.
These bounds are certified because the interval enclosures account for all rounding errors.\par

\section*{S9. Lur'e representation and the discrete Kalman conjecture}

The kicked oscillator can be written as a discrete-time Lur'e system, in which a linear plant is coupled through feedback to a scalar, memoryless nonlinearity. This representation connects the mechanism studied in the Letter with the theory of absolute stability and shows that the oscillator is an explicit counterexample to the discrete-time Kalman conjecture: stability of every constant-gain linear system does not guarantee global stability when the gain varies with the state.

Let $y_*$ be the unique fixed point and set $u_n=y_n-y_*$. The nonlinear feedback can then be written as
\begin{equation}\label{eq:SMpsi}
\psi(u)=\mu[u-\sin(y_*+u)+\sin y_*],
\qquad \psi(0)=0,
\end{equation}
with derivative
\[
0\leq\psi'(u)=\mu[1-\cos(y_*+u)]\leq2\mu.
\]
In terms of this feedback, the recurrence takes the form
\begin{equation}\label{eq:SMlure}
u_{n+1}=\mu u_n-bu_{n-1}-\psi(u_n).
\end{equation}

The role of the feedback is easiest to see by first replacing it with a constant gain, $\psi(u)=ku$, where $0\leq k\leq2\mu$. The resulting linear system has Jacobian
\[
A(k)=\begin{pmatrix}0&1\\-b&\mu-k\end{pmatrix},
\]
whose characteristic polynomial is $z^2-(\mu-k)z+b$. The Schur conditions reduce to $0<b<1$ and $|\mu-k|<1+b$, i.e., $\mu-1-b<k<\mu+1+b$. Since $k\in[0,2\mu]$ implies $|\mu-k|\leq\mu$, every constant-gain system in the prescribed sector is linearly stable whenever $\mu<1+b$. In the language of absolute stability, the Nyquist gain is $k_N=\mu+1+b$, which exceeds the maximal slope $2\mu$ of $\psi$ precisely when $\mu<1+b$.

The nonlinear oscillator is different because its incremental gain, $k_n=\psi'(u_n)$, changes along the trajectory. Although each frozen matrix $A(k_n)$ is stable under the same condition, their ordered products can amplify perturbations exponentially. The discrete-time Kalman conjecture asserts that a Lur'e system whose nonlinearity satisfies $\psi(0)=0$ and $0\leq\psi'\leq k$ is globally asymptotically stable whenever $k<k_N$. Here $\psi$ meets these hypotheses with $k=2\mu<k_N$, yet the fixed point $u=0$ coexists with periodic and chaotic attractors (Letter and Sec.~S7) and is therefore not globally attracting. The kicked oscillator is thus an explicit counterexample, with a smooth nonlinearity and chaotic rather than merely periodic non-convergent dynamics; second-order counterexamples with periodic cycles were constructed in Ref.~\cite{Heath2015}. Stability for every constant feedback gain therefore does not rule out instability under state-dependent nonlinear feedback.

With the feedback sign included, the transfer function of the linear plant is
\begin{equation}\label{eq:SMtransfer}
H(z)=-\frac{z}{z^2-\mu z+b};
\end{equation}
equivalently, Eq.~\eqref{eq:SMlure} is the standard negative-feedback loop $u=-G(z)\psi(u)$ with $G(z)=-H(z)=z/(z^2-\mu z+b)$, a plant of the same second-order form as those considered in Ref.~\cite{Heath2015}.
This form connects the oscillator directly to absolute-stability analyses such as that of Ref.~\cite{Heath2015}. The feedback sector and the stability conditions for constant gains are explicit, yet the nonlinear system supports chaotic dynamics. The joint-spectral-radius bounds and geometric-pumping decomposition developed in the Letter can thus be viewed as quantitative tools for examining how time-dependent feedback gains combine to produce growth that is absent from each constant-gain system considered separately.

\section*{S10. Experimental implementation and tests}

\emph{Proposed experimental implementation.}
A possible realization of the non-normal route to chaos is a laser-cooled trapped ion driven by a pulsed optical standing wave. This proposal builds on trapped-ion friction simulators~\cite{Bylinskii2015}, theoretical schemes for kicked nonlinear oscillators~\cite{Breslin1997}, and real-time monitoring of single-ion motion~\cite{Bushev2013}. By tuning the pulse interval, optical intensity, trap position, and damping, one can aim to reproduce the dissipative return map and test its central prediction: positive Lyapunov growth even though the local Jacobian spectrum remains contracting throughout phase space.

\emph{Single-ion architecture and controls.}
The setup would use one motional degree of freedom of an ion confined in an approximately harmonic trap, with axial frequency $\omega_0$ and a velocity-damping coefficient $2\gamma$ over the relevant range of oscillation amplitudes. An optical standing wave provides a sinusoidal force with spatial wave number $k$. Short, phase-coherent intensity pulses of effective duration $\tau$ are applied at a programmable interval $T$, and the trap minimum $x_0$ is calibrated relative to the standing-wave phase. In the impulsive limit, the dynamics is described by Eq.~\eqref{eq:eom}. The parameters entering Eq.~\eqref{eq:physical} can then be determined from independent measurements of the oscillation frequency, damping, optical phase, pulse impulse, and timing, rather than inferred from the observation of chaos:
\begin{equation}\label{eq:SMexpcontrols}
\begin{gathered}
\omega=(\omega_0^2-\gamma^2)^{1/2},\quad
\theta=\omega T,\quad \lambda=e^{-\gamma T},\\
b=e^{-2\gamma T},\qquad
\mu=e^{-\gamma T}\frac{kF_0\tau}{m\omega},\qquad
C=(1+b)kx_0\quad(\theta=\pi/2).
\end{gathered}
\end{equation}
The standing-wave phase offers an additional means of adjusting the offset relative to the trap. Cooling conditions mainly control $\gamma$, while laser intensity and pulse duration set the impulse $F_0\tau$. In practice, these controls may not be fully independent, so their effects should be calibrated separately.

For the working point $(b,\mu,C)=(0.9,1.7076,1.885)$, a useful set of dimensionless targets is
\begin{equation}\label{eq:SMexpbenchmark}
\gamma T=-\tfrac12\ln(0.9)\simeq0.05268,
\quad K=\frac{\mu}{\sqrt b}\simeq1.80,
\quad kx_0=\frac{C}{1+b}\simeq0.9921.
\end{equation}
With a quarter-period free rotation, $T=\pi/(2\omega)$, this corresponds to $\gamma/\omega=-\ln(0.9)/\pi\simeq0.0335$. Because only $\theta$ modulo $2\pi$ enters the map, the same $(b,\mu,C)$ is obtained with $T=(4\ell+1)\pi/(2\omega)$ for any integer $\ell\geq0$, which lowers the required damping to $\gamma/\omega=-\ln(0.9)/[(4\ell+1)\pi]$ (e.g., $0.0067$ for $\ell=1$ and $0.0037$ for $\ell=2$), at the price of a longer pulse interval and a $(4\ell+1)$-fold larger sensitivity of $\theta$ to drifts of $\omega$. These values define a benchmark rather than an established experimental operating point; the attainable signal-to-noise ratio and calibration stability remain to be assessed for a specific apparatus.

\emph{Conditions for the return-map approximation.}
The impulsive description requires the pulse to be short enough that the ion's position and free harmonic motion change negligibly during the kick. Representative conditions are $\omega_0\tau\ll1$ and $\tau\ll T$. The optical force must remain approximately sinusoidal over the range explored by the ion, since higher spatial harmonics or appreciable changes in the trap confinement would alter the Jacobian. Likewise, the cooling force should be close to linear in velocity over the visited range. Trap micromotion, timing jitter, finite pulse width, intensity noise, photon-recoil diffusion, and measurement backaction may all perturb the deterministic map and should be characterized separately. In particular, if the damping is produced by laser cooling, it is necessarily accompanied by momentum diffusion, which fixes the steady-state rms excursion $x_{\rm th}$ of the cooled ion. Between kicks the motion then relaxes as an Ornstein--Uhlenbeck process, and each interval $T$ adds independent noise of rms $kx_{\rm th}\sqrt{1-b}$ to each of the rescaled coordinates $X$ and $P$. This noise must be small compared with the scale of the structures of the attractor in Fig.~\ref{fig:attractor}(a). For Doppler-limited ions in MHz traps driven by an optical standing wave, $kx_{\rm th}$ can be of order unity, so that sub-Doppler cooling or a damping mechanism with a lower noise temperature may be required; this ratio should be evaluated for the specific apparatus.
A classical trajectory description is appropriate only if the motional state and measurement protocol resolve the relevant phase-space structures without quantum fluctuations or heating overwhelming the kick-to-kick dynamics. A meaningful test therefore requires independent calibration of the forcing, damping, pulse timing, and noise, followed by direct checks of the stroboscopic map and its local Jacobians.

\emph{Discriminating observables.}
Sampling the position immediately before each pulse gives $X_n=k(x_n-x_0)$ and $y_n=kx_n$. At quarter-period timing, this sequence can be compared directly with the two-step recurrence~\eqref{eq:rec}; a separate momentum measurement is not required for this test. Once $b$, $\mu$, and the optical phase have been independently calibrated, the local curvature and the corresponding one-kick Jacobian follow from
\begin{equation}\label{eq:SMexpJac}
t_n=\mu\cos y_n,\qquad
J_n=\begin{pmatrix}0&1\\-b&t_n\end{pmatrix},
\qquad \rho(J_n)=\sqrt b<1.
\end{equation}
The measured sequence can then be used to reconstruct the finite-product rates $h_L$ of Eq.~\eqref{eq:hL} and to integrate the tangent dynamics to estimate $\lambda_1$. These model-based estimates should be compared with independent measurements of the separation of initially neighboring trajectories. Repeated preparations can help distinguish reproducible deterministic stretching from stochastic diffusion. The principal test is positive long-time Lyapunov growth alongside independent confirmation, from the calibrated return map, that the local Jacobian spectral radius is uniformly below unity throughout phase space.

\emph{Mapping coexistence and intermittency.}
At fixed $b$ and $\mu$, the parameter $C$ can be scanned by changing the relative position of the trap and standing wave. For each value, reproducible sets of initial positions and velocities can be prepared and the fractions captured by the stable fixed point, periodic attractors, and the predicted chaotic state recorded. The observation time must be long enough to detect delayed capture from chaotic transients: a positive finite-time Lyapunov exponent alone does not establish permanent attraction. Near $C_f$, laminar episodes can be identified using a specified tolerance on $|y_n-y_{n-24}|$. Their residence times should be recorded before any capture by a periodic orbit, and their dependence on $C-C_f$ fitted over a documented scaling window. This would allow a test of the predicted scaling
\[
\langle\tau_{\rm lam}\rangle\propto
(C-C_f)^{-0.500\pm0.002},
\]
while sufficiently long runs would help distinguish sustained intermittent chaos from long-lived chaotic transients near $C_f$.

\textit{Varying the pulse timing.}
The analysis above assumes that successive kicks are separated by exactly a
quarter of the damped oscillation period, $\theta=\omega T=\pi/2$. In practice,
the pulse interval $T$ or the trap frequency $\omega$ will be slightly off, and
deliberately changing $T$ also provides a useful test. Two questions then arise:
does every local Jacobian remain contracting, and does chaos survive?

The first question has a simple answer. Changing $T$ alters both the rotation
angle $\theta=\omega T$ between kicks and the damping $\gamma T$ accumulated
between them. All local eigenvalues remain inside the unit circle as long as
$2|\cos\theta|+K\sin\theta<2\cosh(\gamma T)$, which follows from the standard
stability condition $|\mathrm{tr}\,J|<1+\det J$ for $2\times2$ matrices. At the
working point ($K=1.8$), this allows $\theta$ to range from $83.9^\circ$ to
$96.1^\circ$, slightly wider than the window $83.97^\circ$--$96.03^\circ$ quoted in
the Letter, within which the eigenvalues are in addition complex conjugates. In
other words, the product $\omega T$ may deviate by up to about $7\%$ from its
quarter-period value; for the longer pulse intervals $T=(4\ell+1)\pi/(2\omega)$ with integer
$\ell\ge 1$ introduced above, the allowed deviation is $4\ell+1$ times smaller.
Inside this window, any chaos observed is still chaos
without local spectral instability. Outside it, some Jacobians have an eigenvalue
larger than one, and the experiment no longer tests the mechanism proposed here.
Independent measurements of $\omega$ and $T$ should therefore confirm that the
timing stays within this window.

For the second question, the computer-assisted proof of chaos (Sec.~S8) does
not rely on fine-tuning: every inequality it checks holds with a finite margin.
The same proof therefore remains valid when $\theta$ is close enough to
$\pi/2$~\cite{ZgliczynskiGidea2004}, so chaos, in the sense of positive topological
entropy, survives small timing errors. How large an error is allowed can be
determined by repeating the verification at $\theta\neq\pi/2$. This result does
not guarantee that trajectories actually settle onto the chaotic attractor: as
$\theta$ changes, the chaotic set could stop attracting nearby states, or more
initial conditions could be captured by the stable fixed point. This is what an
experiment should measure. In practice, one would (i) select several values of
$T$ across the allowed window; (ii) for each value, prepare many initial
positions and velocities and record whether each run settles onto the fixed
point, onto a periodic orbit, or continues to move irregularly; and (iii) for the
irregular runs, estimate the Lyapunov exponent from the recorded positions
$X_n$, as described above. A positive exponent that persists across the window,
together with a nonzero fraction of irregular runs, would show that the
mechanism does not depend on exact quarter-period timing.

\section*{S11. Mechanical energy input and tangent-space amplification}
\enlargethispage{4\baselineskip}

The work done by the pulses and the growth of infinitesimal
perturbations describe two distinct balances. The mechanical
energy of the oscillator is
\begin{equation}\label{eq:SMenergy}
\mathcal E(t)=\tfrac12m\dot x^2+
\tfrac12m\omega_0^2(x-x_0)^2.
\end{equation}
Between kicks, Eq.~\eqref{eq:eom} gives
\begin{equation}\label{eq:SMdissipation}
\frac{d\mathcal E}{dt}=-2m\gamma\dot x^2\le0.
\end{equation}
At each kick the position remains continuous, while the velocity
changes by $\Delta v=(F_0\tau/m)\sin(kx)$. If $\dot x^-$ denotes
the velocity immediately before the pulse, the work associated
with the impulse is exactly
\begin{equation}\label{eq:SMwork}
\Delta\mathcal E
=m\dot x^-\Delta v+\tfrac12m(\Delta v)^2.
\end{equation}
Depending on the phase and the incoming velocity, a pulse can
therefore either add or remove mechanical energy. In statistically
stationary motion, the mean work per kick balances the energy lost
to viscous dissipation during the following free evolution:
\begin{equation}\label{eq:SMstationarywork}
\left\langle\Delta\mathcal E_n\right\rangle
=2m\gamma\left\langle
\int_{nT^+}^{(n+1)T^-}\dot x(t)^2\,dt
\right\rangle.
\end{equation}
This relation concerns the energy of the particle's actual
trajectory and the work performed by the external drive.

The identity in Eq.~\eqref{eq:pump} describes something different.
It applies to an infinitesimal perturbation $\xi_n$ and uses the
quadratic form $E_n=\xi_n^\mathsf T G(t_n)\xi_n$, whose metric
varies with the local substrate curvature. This metric is introduced
to describe contraction in tangent space; its relation to the mechanical energy is made explicit by Eq.~\eqref{eq:SMphys}. In the rescaled variables, $(2k^2/m)\,\mathcal E=\omega^2(X^2+P^2)-2\gamma\omega XP+2\gamma^2X^2$, so that the free part $\delta X_n^2+\delta P_n^2$ of $E_n$ coincides with the Hessian form of $\mathcal E$ up to relative corrections of order $\gamma/\omega\simeq0.03$ at the working point. The essential difference is the cross term $K\cos y_n\,\delta X_n\delta P_n$, which tilts the metric according to the substrate curvature at the current kick and has no counterpart in $\mathcal E$. The term
$W_n=-(t_{n+1}-t_n)u_{n+1}v_{n+1}$ records the change associated
with passing between successive local contraction metrics. It is
not the work $\Delta\mathcal E_n$ supplied by a pulse. In the oscillator variables, it equals the change of kick shear, $K(\cos y_{n+1}-\cos y_n)$, times the correlation $\delta X_{n+1}\delta P_{n+1}$ of the perturbation [Eq.~\eqref{eq:SMphys}], a purely geometric quantity. Its logarithmic
contribution to Eq.~\eqref{eq:lambda} can outweigh the uniform
contraction $\tfrac12\ln b<0$, even when the mean mechanical
power input does no more than compensate for viscous dissipation.

The two balances are therefore consistent: the drive sustains the
bounded motion of the particle, while the sequence of substrate
curvatures visited along that motion determines whether nearby
trajectories approach one another or separate. The resulting growth
comes from the ordered composition of non-normal Jacobians, not from
any creation of mechanical energy by non-normality. For the numerical
attractor, the geometric contribution in Eq.~\eqref{eq:lambda} is
$0.1557\pm0.0002$. It exceeds the magnitude of the fixed spectral
contraction, $-\tfrac12\ln b\simeq0.0527$, and gives the measured
maximal Lyapunov exponent $\lambda_1\simeq0.103$; numerical checks
are reported in Sec.~S7.


\begin{thebibliography}{99}

\bibitem{Lyapunov1892} A. M. Lyapunov, \emph{The General Problem of the Stability of Motion} (Taylor \& Francis, London, 1992), translated by A. T. Fuller from the Russian doctoral thesis (Kharkov Mathematical Society, Kharkov, 1892).

\bibitem{Trefethen1993} L. N. Trefethen, A. E. Trefethen, S. C. Reddy, and T. A. Driscoll, Science \textbf{261}, 578 (1993).

\bibitem{Sornette2026PRE} D. Sornette, V. R. Saiprasad, and V. Troude, Phys. Rev. E \textbf{114}, 024222 (2026).

\bibitem{Sornette2026arXiv} D. Sornette, V. R. Saiprasad, and V. Troude, arXiv:2609.18017 (2026).

\bibitem{CimaGasullManosas1999} A. Cima, A. Gasull, and F. Ma\~nosas, Nonlinear Anal. \textbf{35}, 343 (1999).

\bibitem{AddasGomes2011} S. Addas-Zanata and B. Gomes, Qual. Theory Dyn. Syst. \textbf{10}, 327 (2011), doi:10.1007/s12346-011-0043-z.

\bibitem{Prandtl1928} L. Prandtl, Z. Angew. Math. Mech. \textbf{8}, 85 (1928).

\bibitem{Tomlinson1929} G. A. Tomlinson, Philos. Mag. \textbf{7}, 905 (1929).

\bibitem{Vanossi2013} A. Vanossi, N. Manini, M. Urbakh, S. Zapperi, and E. Tosatti, Rev. Mod. Phys. \textbf{85}, 529 (2013).

\bibitem{Bylinskii2015} A. Bylinskii, D. Gangloff, and V. Vuleti\'c, Science \textbf{348}, 1115 (2015).

\bibitem{TroudeSornette2026} V.~Troude and D.~Sornette, Phys. Rev. Research \textbf{8}, 033050 (2026).

\bibitem{RotaStrang1960} G.-C. Rota and G. Strang, Indag. Math. \textbf{22}, 379 (1960).

\bibitem{Jungers2009} R. M. Jungers, \emph{The Joint Spectral Radius: Theory and Applications} (Springer, Berlin, 2009).

\bibitem{ZgliczynskiGidea2004} P. Zgliczy\'nski and M. Gidea, J. Differ. Equations \textbf{202}, 32 (2004).

\bibitem{Ruelle1978} D. Ruelle, Bol. Soc. Bras. Mat. \textbf{9}, 83 (1978).

\bibitem{PomeauManneville1980} Y. Pomeau and P. Manneville, Commun. Math. Phys. \textbf{74}, 189 (1980).

\bibitem{Heath2015} W. P. Heath, J. Carrasco, and M. de la Sen, Automatica \textbf{60}, 140 (2015).

\bibitem{Breslin1997}
J. K. Breslin, C. A. Holmes, and G. J. Milburn, Phys. Rev. A \textbf{56}, 3022 (1997).

\bibitem{Bushev2013}
P. Bushev, G. H\'etet, L. Slodi\v{c}ka, D. Rotter, M. A. Wilson, F. Schmidt-Kaler, J. Eschner, and R. Blatt, Phys. Rev. Lett. \textbf{110}, 133602 (2013).

\end{thebibliography}
\end{document}